\documentclass[aps,prb,twocolumn,longbibliography,nofootinbib,floatfix,superscriptaddress]{revtex4-2}
\usepackage{physics}
\usepackage{graphicx}
\usepackage{dcolumn}

\usepackage{subfigure}
\usepackage{dsfont}
\usepackage{capt-of}

\usepackage[english]{babel}
\usepackage[utf8]{inputenc}
\usepackage[T1]{fontenc}
\usepackage{mathrsfs}
\usepackage{amsfonts}
\usepackage{bbm}
\usepackage{bm}
\usepackage{enumerate}
\usepackage[dvipsnames]{xcolor}
\usepackage{float}
\usepackage{subdepth}
\usepackage{fnpct}
\usepackage{booktabs}
\usepackage{multirow}
\usepackage[math]{cellspace}

\usepackage{lipsum}
\usepackage{comment}
\usepackage{soul}
\usepackage[normalem]{ulem}

\usepackage[colorlinks,
linkcolor=BrickRed,
citecolor=MidnightBlue,
urlcolor=MidnightBlue,
bookmarks=true,        
bookmarksopen=true,    
bookmarksnumbered=true,
]{hyperref}

\newcounter{ls}

\newcommand{\pone}{\bs{+^{14;23}}}

\newcommand{\pthree}{\bs{+^{1234}}}
\newcommand{\bu}{\bs{\mathcal{U}}}
\newcommand{\bk}{\bs{\mathcal{K}}}
\begin{document}

\begin{minipage}[t]{1\columnwidth}%

\global\long\def\kket#1{\left\Vert #1\right\rangle }%

\global\long\def\bbra#1{\left\langle #1\right\Vert }%

\global\long\def\bbrakket#1#2{\left\langle #1\right. \left\Vert #2\right\rangle }%

\global\long\def\av#1{\left\langle #1 \right\rangle }%

\global\long\def\tr{\text{tr}}%

\global\long\def\Tr{\text{Tr}}%

\global\long\def\im{\text{Im}}%

\global\long\def\re{\text{Re}}%

\global\long\def\sgn{\text{sgn}}%

\global\long\def\Det{\text{Det}}%

\global\long\def\pf{\text{pf}}%

\global\long\def\abs#1{\left|#1\right|}%

\global\long\def\up{\uparrow}%

\global\long\def\down{\downarrow}%

\global\long\def\E{\mathrm{e}}%

\global\long\def\I{\mathrm{i}}%

\global\long\def\vc#1{\mathbf{#1}}%

\global\long\def\rm#1{\boldsymbol{#1}}%

\global\long\def\bs#1{\boldsymbol{#1}}%

\global\long\def\t#1{\text{#1}}%
\end{minipage}

\title{Dissipation-Induced Threshold on Integrability Footprints }

\author{Rodrigo M. C. Pereira}\email{rodrigo.m.c.pereira.6@tecnico.ulisboa.pt}
\affiliation{CeFEMA-LaPMET, Physics Department, Instituto Superior Técnico, Universidade de Lisboa, 1049-001 Lisboa, Portugal}
\affiliation{Institute of Computer Architecture and Computer Engineering, University of Stuttgart, Germany}
\affiliation{Center for Integrated Quantum Science and Technology, Germany}

\author{Nadir Samos S{\'a}enz de Buruaga }
\affiliation{CeFEMA-LaPMET, Physics Department, Instituto Superior Técnico, Universidade de Lisboa, 1049-001 Lisboa, Portugal}
\affiliation{CIAF,Departamento de Física Teórica, Universidad Aut\'onoma de Madrid, C. Francisco Tom\'as y Valiente 7, 28049 Madrid, Spain}

\author{Kristian Wold}
\affiliation{Department of Computer Science, OsloMet – Oslo Metropolitan University, 0130 Oslo, Norway}

\author{Heine Olsson Aab{\o}}
\affiliation{Department of Computer Science, OsloMet – Oslo Metropolitan University, 0130 Oslo, Norway}

\author{Lucas Sá}
\affiliation{TCM Group, Cavendish Laboratory, University of Cambridge, JJ Thomson Avenue, Cambridge CB3 0HE, UK\looseness=-1}

\author{Sergey Denisov}
\affiliation{Department of Computer Science, OsloMet – Oslo Metropolitan University, 0130 Oslo, Norway}
\affiliation{Norwegian Quantum Software Centre (NorQSoft), Simula Lab, Kristian Augusts gate 23, 0164 Oslo}

\author{Pedro Ribeiro}\email{ribeiro.pedro@tecnico.ulisboa.pt}
\affiliation{CeFEMA-LaPMET, Physics Department, Instituto Superior Técnico, Universidade de Lisboa, 1049-001 Lisboa, Portugal}
\affiliation{Beijing Computational Science Research Center, Beijing 100193, China}

\begin{abstract}
The presence of a dissipative environment substantially disrupts unitary evolution. Nevertheless, key features of the evolution, such as its integrable or chaotic nature, are not immediately erased by dissipation. To understand this, we model environment-induced dissipation as a convex combination of unitary evolution and a random Kraus map, and examine how signatures of integrability gradually fade with increasing dissipation strength. Our analysis shows that in the weakly dissipative regime, the complex eigenvalue spectrum organizes into well-defined, high-density clusters. We estimate the critical dissipation threshold beyond which these clusters disappear, rendering the dynamics indistinguishable from chaotic evolution. This threshold depends only on the number of spectral clusters and the rank of the Kraus map. To characterize this transition, we introduce the eigenvalue angular velocity as a diagnostic of integrability loss. We illustrate our findings using several integrable quantum circuits, including the dissipative quantum Fourier transform. Our results provide a quantitative picture of how noise gradually erases the footprints of unitary integrability in open quantum systems.
\end{abstract}

\maketitle

\textit{Introduction.---}%
The study of quantum chaos has gained renewed attention due to its relevance in various areas, including non-equilibrium dynamics of many-body quantum systems and quantum information processing~\cite{kobrin2021many, lantagne2020diagnosing, sieberer2019digital, rath2021quantum}.
\par
In closed quantum systems, spectral analysis is widely used to identify chaotic behavior~\cite{haake1991quantum}. A key conjecture states that the level statistics of integrable systems follow a Poisson distribution~\cite{level_stats}, while chaotic systems exhibit level statistics described by random matrix theory~\cite{level_stats2}. The hallmark of integrable dynamics is level clustering, where the probability density is highest at zero spacing. In contrast, chaotic systems display level repulsion, with the probability density vanishing at small spacings.  Even though there is not yet rigorous proof, these statements are strongly supported by several numerical studies~\cite{haake1991quantum}. It has also been shown that even small perturbations can induce chaos in the thermodynamic limit, thus breaking integrability~\cite{rabson2004crossover}.
\par
Sampling random unitaries according to the Haar measure on the unitary group, that is, from the circular unitary ensemble (CUE), reliably leads to chaotic spectral statistics. Meanwhile, many physically relevant models remain integrable, such as classical spin chains and free-fermion systems. In the context of quantum information, the latter are also known as matchgate circuits—quantum circuits that emulate the dynamics of non-interacting fermions and can be efficiently simulated on a classical computer~\cite{valiant2001quantum, terhal2002classical, jozsa2008matchgates}. Another highly structured gate set that lacks universal dynamics is the Clifford group. Clifford circuits can be simulated in polynomial time using the stabilizer formalism~\cite{gottesman1998heisenberg, aaronson2004improved}, making them useful in quantum many-body physics and the benchmark of quantum computers~\cite{li2018quantum, magesan2011scalable, onorati2019randomized, helsen2022general}.
\par
Since no real quantum system is perfectly isolated, understanding how chaos and integrability manifest in open quantum systems is essential. In open systems, evolution is governed by non-unitary operators, leading to decoherence and dissipation as information is lost to the environment. Despite this, spectral statistics remain a powerful tool for diagnosing chaotic behavior. Level spacing distributions provide insight into the nature of the system: integrable systems exhibit statistics described by a two-dimensional Poisson process, whereas chaotic systems follow the Ginibre unitary ensemble (GinUE) distribution~\cite{ginibre1965,Haake_spacing, akemann_spacings}. Other measures, such as complex spacing ratios (CSRs)~\cite{sa2020complex,garcia2022PRX,rubio2022integrability, sa2021integrable, shivam2023many}, further characterize this distinction: integrable systems yield uniformly distributed points on the unit circle, whereas chaotic systems form a distinctive ``bitten-doughnut'' shape.
\par
Dissipation is expected to impact integrable unitary dynamics, eventually leading to universal dissipative behavior. However, the extent to which integrable features persist under generic dissipation channels remains an open question.
In this work, we seek to quantify how hallmarks of integrability---such as level clustering---gradually fade as dissipation increases. We model the open-system dynamics as a convex combination of the original unitary map and a random Kraus operator, with the rank of the operator corresponding to the number of dissipative channels. 
Interestingly, while conventional measures such as quantum fidelity fail to detect the erosion of integrable behavior, spectral diagnostics offer a more sensitive and intuitive probe. To formalize this spectral signature, we introduce the eigenvalue angular velocity, which captures the impact of dissipation on integrable dynamics. Our results show that level clustering, a defining characteristic of integrability, remains robust against weak dissipation.
\par
The main contribution of this study is the identification of a universal relationship governing the critical dissipation strength beyond which spectral clustering disappears, for which we derive an analytical expression that accurately approximates the numerical results. Within the diluted-unitary framework studied here, this critical scale is determined only by two coarse spectral parameters: the number of clusters in the underlying unitary map and the rank of the dissipative channel. It is therefore independent of microscopic details of the noise realization, establishing a universal transition mechanism in this class of open quantum maps. Moreover, the fact that the same framework reproduces key spectral features of more generic dissipative dynamics \cite{wold2024spectra} suggests that this universal scale provides a physically relevant benchmark for the destruction of integrable spectral structure in generic noisy quantum systems.
\par
\textit{Diluted unitary.---}%
We consider a quantum map that deviates from an ideal unitary implementation due to random interactions with the environment. Following the approach in Ref.~\cite{sa2020spectral}, we define
\begin{equation}\label{eq: five maps}
\Phi_{M} = (1-\kappa)U_{M} \otimes U_{M}^* + \kappa \sum_{j=1}^r K_j \otimes K_j^*,
\end{equation}
where $\kappa \in [0,1]$ controls the dissipation strength. When $\kappa = 0$, the evolution is purely unitary, whereas $\kappa = 1$ corresponds to a random rank-$r$ Kraus map, whose operators satisfy $\sum_{j=1}^r K_j^{\dagger}K_j = I$. Here, $U_M$ and $K_j$ are $d \times d$ matrices, the rank of the Kraus operators is constrained by $1 \leq r \leq d^2 - 1$, and the subscript $M$ on $U_M$ denotes the ensemble from which the unitary is sampled. We refer to $\Phi_M$ as a \emph{diluted unitary}.
\par
We aim to investigate the impact of dissipation on various maps exhibiting chaotic and distinct regular dynamics, as well as to identify measures that distinguish their behaviors and responses to the loss of unitarity. Notably, the differences between the models arise solely from the choice of the unitary operator; 
the Kraus map that mimics the complex noisy environment is sampled by generating a large $rd\times rd$ unitary random matrix $W\in\text{CUE}$ and chopping the first $d$ columns into $r$ blocks \cite{bruzda2009random}, which we take to be the Kraus operators, $(K_j)_{ab}=W_{(j-1)d+a,b}$, with $a,b=1,\dots,d$, $j=1,\dots,r$.
The unitarity of $W$ ensures that the constraint on the Kraus operators is satisfied. 
\par
We analyze a chaotic case, with $U_M$ taken to be a random matrix sampled from the CUE, and three integrable models: (i) an operator from the Clifford group sampled uniformly following Ref.~\cite{van2021simple}, (ii) an integrable periodically driven (or Floquet) spin-1/2 chain based on Ref.~\cite{vanicat2018integrable}, and (iii) a random matrix implementing free-fermion dynamics, i.e., a random matchgate circuit.
These systems are defined on $L$ sites, such that the Hilbert space dimension is $d=2^L$.
A detailed description of $U_M$ and the sampling procedure for all cases can be found in Supplemental Material (SM)~\ref{app: methods}.

\textit{Integrability footprints.---}%
As a first approach, we analyze the quantum fidelity between a clean pure state $|\Psi\rangle$ and its actual noisy realization $\rho$, $\mathcal{F} = \mel{\Psi}{\rho}{\Psi}$, for a random quantum circuit. Each layer comprises a random map of the form Eq.~\eqref{eq: five maps} for a specific choice of ensemble and constant dissipation strength $\kappa$. Each time step, we sample a new random diluted unitary.
Since the Kraus operators are obtained from a Haar-random matrix, they inherit the left-right invariance property. Consequently, the average fidelity is fully determined by the statistics of the Kraus operators, and the specific unitary ensemble becomes irrelevant. Therefore, this measure is not able to distinguish the different models, and therefore, the loss of integrability of the systems, as discussed in SM~\ref{app: fidelity}.

\begin{figure}[tbp]
    \includegraphics[width=\linewidth]{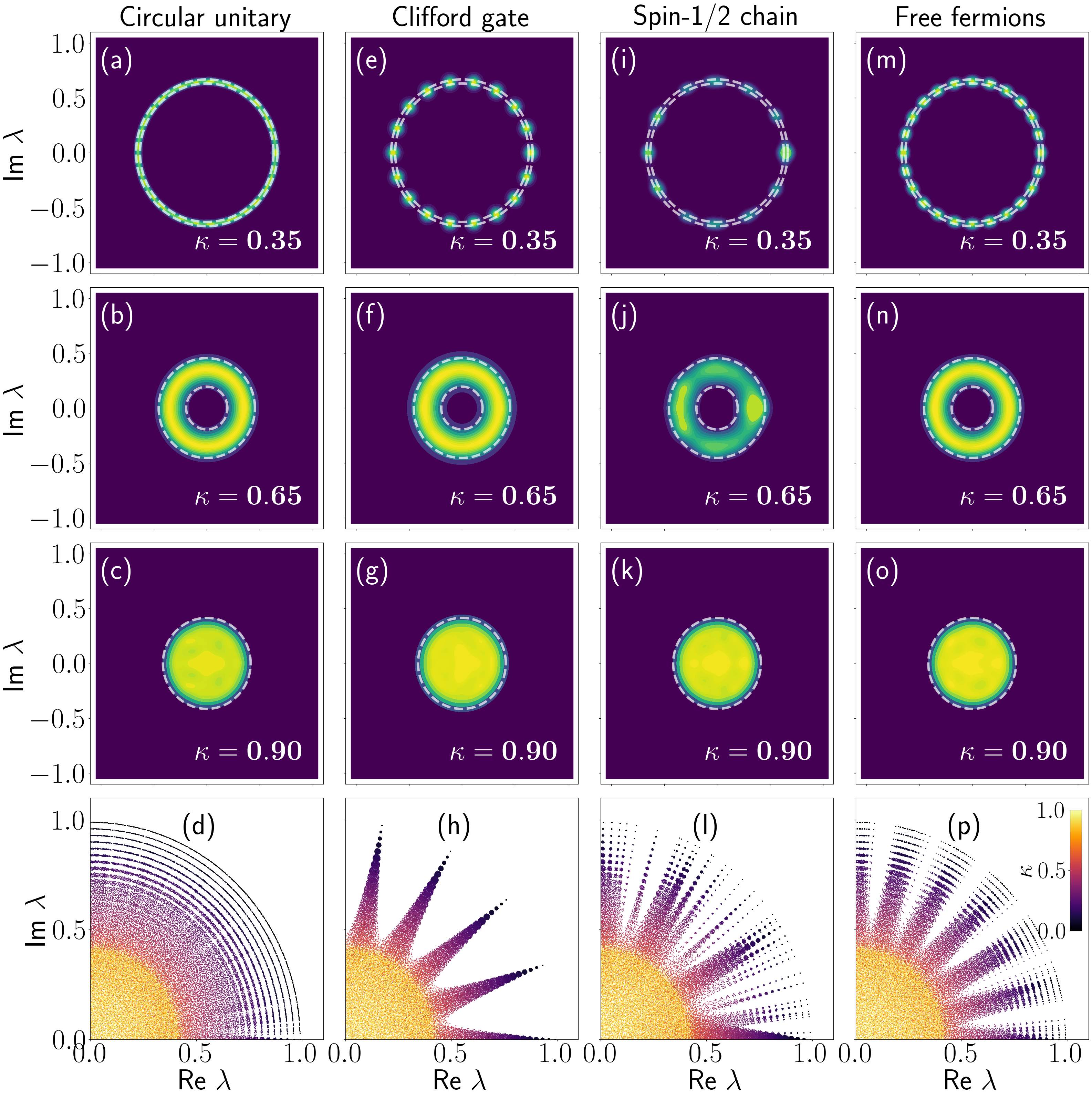}
    \caption{
    Spectrum of the quantum maps in Eq.~\eqref{eq: five maps} for $L = 6$ and $r = 5$. Each column represents a different model based on the ensemble of the unitary circuit, indicated on top. The first three rows depict the eigenvalue density for the indicated value of $\kappa$. The white dashed lines are the circumferences with inner and outer radii given by Eq.~\eqref{eq: radii}. The spectrum of the chaotic model in the first column is characterized by a single R-D transition, whereas the integrable models in the remaining columns have an additional C-R transition. The last row depicts the trajectories of the eigenvalues with increasing $\kappa$. For clarity, only the positive quadrant of the complex plane is displayed.}
    \label{fig: spectrum}
\end{figure}

We now turn to the spectral properties of the maps. Figure~\ref{fig: spectrum} presents the spectra of the four models introduced previously.
Each column corresponds to a different model, with the first three rows displaying the eigenvalue distributions for a given dissipation strength $\kappa$, and the last row illustrating the eigenvalue trajectories in the complex plane as $\kappa$ increases. 
While there is always at least one stationary state with eigenvalue 1, its contribution to the eigenvalue density is negligible in the large-system-size limit, so it is not represented.

\par
The chaotic model in the first column was previously analyzed in Ref.~\cite{sa2020spectral} and is characterized by a uniform angular distribution of eigenvalues within a ring of outer ($R_+$) and inner ($R_-$) radii:
\begin{equation}\label{eq: radii} 
R_{\pm} = \frac{1}{\sqrt{r}} \sqrt{(1-\kappa)^2 r \pm \kappa^2}. 
\end{equation}
This behavior is illustrated in Figs.~\ref{fig: spectrum} \textbf{(a)} and \textbf{(b)}. At the critical value
\begin{equation} \label{eq:kappa_R-D}
\kappa_{\text{R-D}} = \frac{1}{1 + 1/\sqrt{r}},
\end{equation}
where $R_- = 0$, a ring-to-disk (R-D) transition occurs, as shown in Fig.~\ref{fig: spectrum} \textbf{(c)}. At the limit $\kappa=1$, the spectral support becomes a disk of radius $1/\sqrt{r}$, which is typical of completely random full-rank maps~\cite{bruzda2009random}.

\par
When considering a unitary operation with degeneracies, as is typical for regular dynamics due to level clustering, there are areas with a significantly high eigenvalue density and regions where it vanishes. We call these structures \emph{clusters}, which can be seen in Figs.~\ref{fig: spectrum} \textbf{(e)}, \textbf{(i)} and \textbf{(m)}. These clusters persist up to a critical value $\kappa_{\text{C-R}}$. Beyond this threshold, the initial structure of the unitary map is no longer discernible, and a cluster-to-ring (C-R) transition occurs, as illustrated in Figs.~\ref{fig: spectrum} \textbf{(f)}, \textbf{(j)} and \textbf{(n)}. The resulting ring has approximately the same inner and outer radii given by Eq.~\eqref{eq: radii}. Similarly to the chaotic case, once the inner radius of the ring vanishes, an R-D transition takes place at the same critical value $\kappa_{\text{R-D}}$, as shown in Figs.~\ref{fig: spectrum} \textbf{(g)}, \textbf{(k)} and \textbf{(o)}. This transition is discussed further in SM~\ref{app: ring to disk}.
\par
The eigenvalue trajectories of the chaotic model in Fig.~\ref{fig: spectrum} \textbf{(d)} suggest that, for small values of $\kappa$, eigenvalues exhibit predominantly radial motion due to level repulsion, with no significant angular displacement. However, for the integrable maps in Figs.~\ref{fig: spectrum} \textbf{(h)}, \textbf{(l)} and \textbf{(p)}, this is not the case, as dissipation lifts degeneracies and induces angular variations.
\par
\textit{Eigenvalue angular velocity.---}%
Quantifying angular variation is thus essential for distinguishing chaotic from integrable systems. Given an eigenvalue at a certain dissipation strength $\kappa$, expressed as $\lambda(\kappa) = s(\kappa) e^{i\theta(\kappa)}$, we define its angular velocity as
\begin{equation}\label{eq: angular velocity}
v_{\theta}(\kappa) \equiv \left| \frac{d\theta(\kappa)}{d\kappa}\right| \approx \left| \frac{\theta(\kappa + \Delta\kappa) - \theta(\kappa)}{\Delta\kappa} \right|.
\end{equation}
To describe the overall spectral behavior, we consider the mean angular velocity, $\overline{v_{\theta}}$, averaged over all eigenvalues. In the small $\kappa$ limit, this quantity is expected to vanish for chaotic unitaries but remain finite for integrable ones. As $\kappa$ increases, the clusters of eigenvalues expand and merge, eventually forming a ring. At this stage, the original nature of the unitary operator is no longer discernible. Consequently, for $\kappa>\kappa_{\text{C-R}}$, chaotic and integrable models should exhibit similar angular velocities.
\begin{figure}[tbp]
    \includegraphics[width=\linewidth]{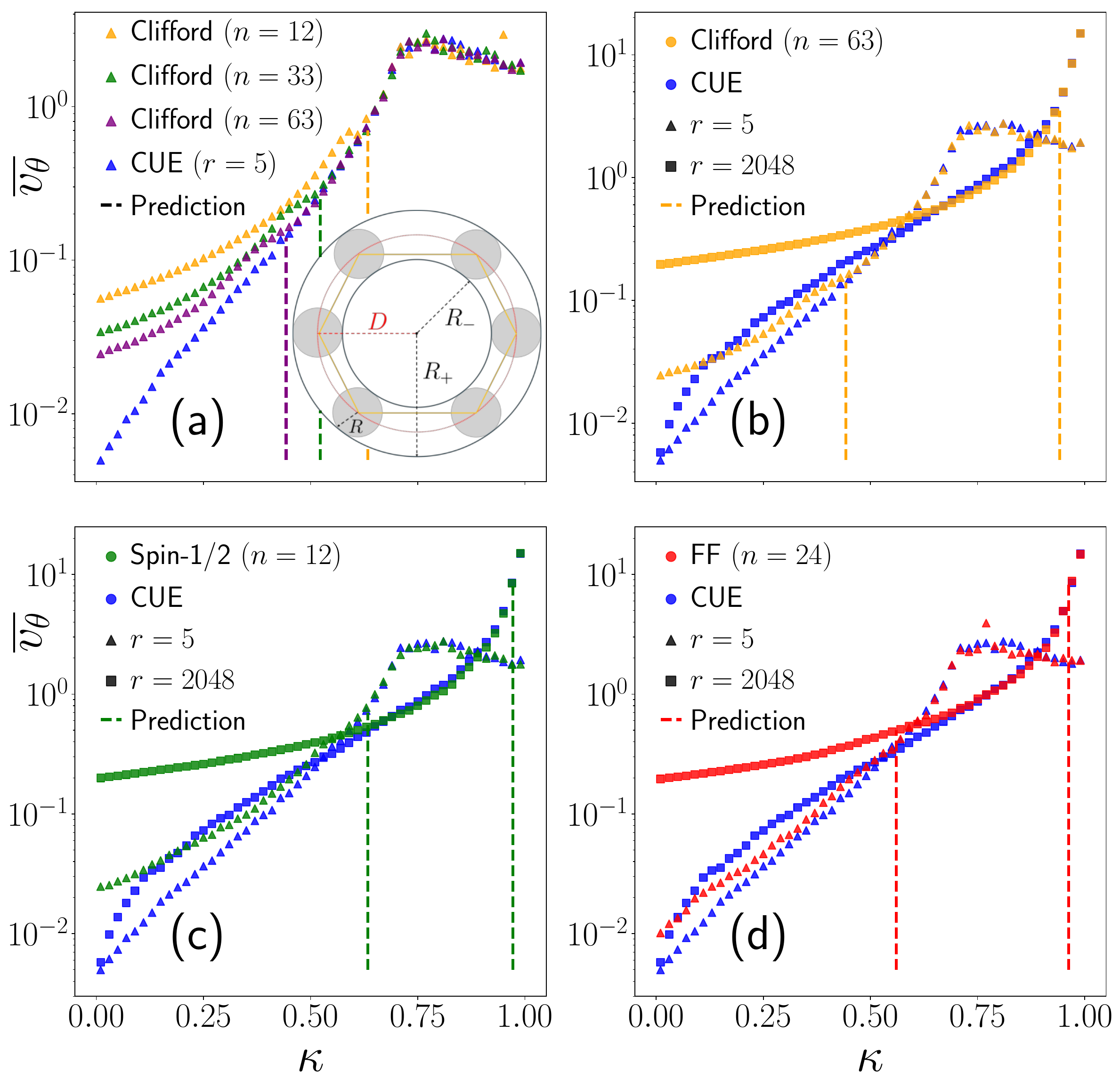}
    \caption{
    Mean eigenvalue angular velocity as a function of dissipation strength, for distinct models and $L = 6$. The C-R transition occurs at the critical value for which the curve of an integrable model starts overlapping with the chaotic one. The dashed lines correspond to the estimate of the transition value $\kappa_{\text{C-R}}$ given by Eq.~\eqref{eq: transition solution}. \textbf{(a)}: CUE model and random Clifford operators with different number of clusters $n$, both with $r = 5$. \textbf{(b)}--\textbf{(d)}: CUE model and the Clifford \textbf{(b)}, integrable spin-1/2 \textbf{(c)}, and free-fermion \textbf{(d)} models, for $r = 5$ and $2048$.
    The inset of \textbf{(a)} depicts the geometric scheme for the derivation of Eq.~\eqref{eq: transition equation}.}
    \label{fig: angular velocity}
\end{figure}
\par
Before numerically analyzing the angular velocity, we derive an estimate for the critical value $\kappa_{\text{C-R}}$, which we identify as the value of dissipation for which clusters occupy the same area as the ring. Consider the idealized case where each cluster forms a perfect circle of radius $R$, with centers arranged in a regular polygon with $n$ vertices (corresponding to the number of clusters) at a distance $D$ from the origin. This geometric configuration is illustrated in the inset of Fig.~\ref{fig: angular velocity} \textbf{(a)} for $n = 6$. Near the C-R transition, the clusters can be assumed within a ring characterized by $R = (R_+ - R_-)/2$ and $D = (R_+ + R_-)/2$, with their centers in the middle of the ring.
The total area covered by the ring is $\pi(R_+^2-R_-^2) = 2\pi\kappa^2/r$, whereas the total area occupied by the clusters is $n\pi R^2$. The transition occurs when these two areas are the same, leading to the condition
\begin{equation}\label{eq: transition equation}
\frac{2\kappa^2}{r} = n\frac{(R_+-R_-)^2}{4}.
\end{equation}
Solving the equation provides an estimate for $\kappa_{\text{C-R}}$, given by 
\begin{equation}\label{eq: transition solution}
    \kappa_{\text{C-R}} = \dfrac{1}{1 + \dfrac{f(n)}{\sqrt{r}}},
\end{equation}
where $f(n) = \sqrt{2/n + n/8}$. This expression reduces to Eq.~\eqref{eq:kappa_R-D} in the case where $n=4$. For sufficiently large $n$, $f(n) \approx \sqrt{n/8}$.
\par
The transition value mainly depends on the ratio $n/r$ and exhibits two limiting behaviors. When $n \gg r$, the transition occurs at smaller values of $\kappa$ as the number of clusters increases, meaning that $\kappa_{\text{C-R}}< \kappa_{\text{R-D}}$. In contrast, when $n \ll r$, the transition value approaches unity, coinciding with the R-D transition, i.e., $\kappa_{\text{C-R}}\to \kappa_{\text{R-D}}$.
\par
It is worth highlighting a somewhat counterintuitive observation: dilute-unitary maps that model environments with many dissipative channels (corresponding to a higher rank $r$) can preserve unitary spectral features under stronger dissipation compared to their low-rank counterparts. In particular, increasing $r$ delays the R–D transition to higher values of the dissipation strength $\kappa$ [see Eq.~\eqref{eq:kappa_R-D}]. Our results further show that higher-rank maps also sustain integrable signatures up to larger values of $\kappa$, whereas concentrating dissipation into a small number of channels proves to be more detrimental to integrable spectral features.
\par
To compute the average angular velocity numerically, we construct the quantum map defined in Eq.~\eqref{eq: five maps} for two close values of $\kappa$ separated by $\Delta\kappa = \kappa/1000$. By identifying the nearest-neighbor eigenvalues between the two maps, we track their trajectories and compute the angular velocity for each eigenvalue using Eq.~\eqref{eq: angular velocity}, averaging over the entire spectrum. To improve numerical stability, real eigenvalues were discarded since their nearest neighbors were often misidentified, leading to abrupt fluctuations in velocity. 
\par
To complete the analysis, we require an operational estimate of the number of spectral clusters $n$. For Clifford circuits, exact degeneracies allow $n$ to be obtained exactly as the number of distinct eigenvalues. For the remaining models, we estimate $n$ from the angular density of the spectrum: after expressing the eigenvalues in terms of angles $\theta_k$, we associate a Gaussian kernel with each $\theta_k$ and count the stable number of local maxima of the resulting density as the smoothing parameter is varied. For the spectra considered here this procedure is robust, because the relevant cluster structure appears as a clear plateau in the cluster count. The method therefore provides a reliable estimate of the quantity that controls Eq.~\eqref{eq: transition solution}. Further details are given in SM~\ref{app: cluster count}, and the larger-system-size data discussed in the End Matter show that the extracted $n$ captures the relevant scaling of the transition.
\par
Figure~\ref{fig: angular velocity} \textbf{(a)} presents numerical results for the Clifford model with different numbers of clusters $n$ and fixed rank $r = 5$. 
As a different number of clusters may arise from randomly sampling Clifford circuits with the same size and depth, we postselect one that yields exactly $n$ clusters. 
Beyond a threshold $\kappa_{\text{C-R}}$, the angular velocities 
converge to that of the chaotic model, indicating that $\overline{v_{\theta}}$ successfully signals the C-R transition. Moreover, systems with fewer clusters, meaning a higher degree of degeneracy, exhibit larger $\kappa_{\text{C-R}}$, in agreement with Eq.~\eqref{eq: transition solution}.
\par
Figures~\ref{fig: angular velocity} \textbf{(b)}--\textbf{(d)} depict the average angular velocity for the Clifford, integrable spin-1/2, and free-fermion models, respectively, for a fixed number of clusters and two different values of the rank $r$.
The results confirm that in the large-$r$ limit, the transition value approaches unity, further supporting the evidence that increasing the rank preserves integrability up to higher values of $\kappa$.

\begin{figure}[tbp]
    \includegraphics[width=\linewidth]{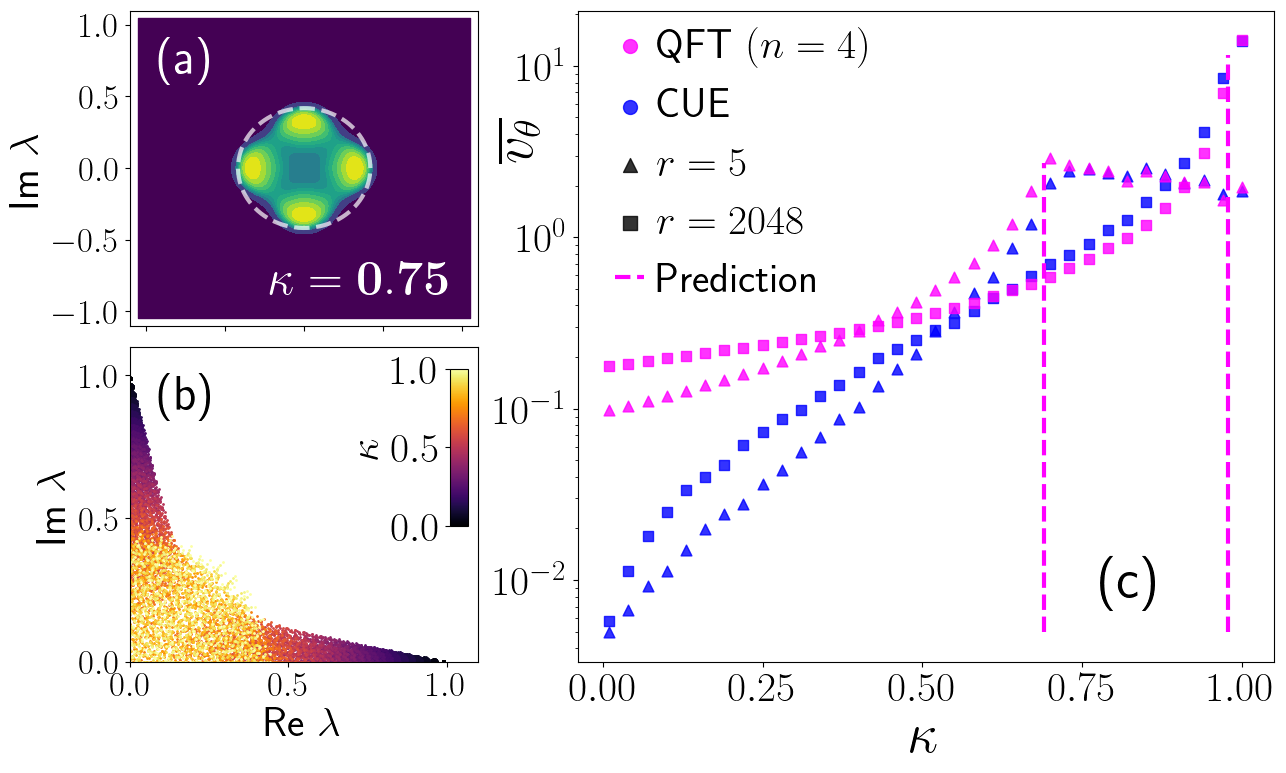}
    \caption{\textbf{(a)}: Eigenvalue density  for $U_M=U_{\text{QFT}}$ with $L = 6$, $r = 5$, and $\kappa = 0.75$. \textbf{(b)}: Eigenvalue trajectories with increasing $\kappa$ for fixed $r = 5$. \textbf{(c)}: Mean eigenvalue angular velocity for the CUE and QFT models for $r = 5$ and $2048$. The dashed lines correspond to the estimate of the transition value $\kappa_{\text{C-R}}$ given by Eq.~\eqref{eq: transition solution}.}
    \label{fig: QFT}
\end{figure}

\textit{Special case of the QFT.---}%
An example of an integrable unitary with a highly structured spectrum is the quantum Fourier transform (QFT) \cite{coppersmith2002approximate}, an algorithm that can be efficiently implemented in a quantum computer. For an $L$-qubit system, the matrix elements of its unitary operator are given by $\left(U_{\text{QFT}}\right)_{jk} = 1/\sqrt{d}\; e^{2\pi ijk/d}$, recalling $d = 2^L$. 
For all system sizes, this operator has a fixed number $n=4$ of distinct eigenvalues, $\pm 1, \pm i$.
\par
In Fig.~\ref{fig: QFT} \textbf{(a)}, we depict the spectrum of the diluted unitary map with $U_M=U_{\text{QFT}}$, for $\kappa = 0.75$ and $r = 5$. Figure~\ref{fig: QFT} \textbf{(b)} shows the trajectories of the eigenvalues as dissipation increases, leading them to converge toward the center until they form a disk. Because the number of clusters is small, distinguishing two separate transitions becomes challenging. Instead, this model appears to exhibit a single cluster-to-disk (C-D) transition, as $\kappa_\text{C-R}$ in Eq.~\eqref{eq: transition solution} reduces to $\kappa_\text{R-D}$ in Eq.~\eqref{eq:kappa_R-D} for $n = 4$. It is worth noting that, according to Eq.~\eqref{eq:kappa_R-D}, the R-D transition is expected to occur at $\kappa_{\text{R-D}} \approx 0.69$ for $r = 5$. However, Fig.~\ref{fig: QFT} \textbf{(a)} shows that the disk still does not homogeneously cover the entire circle.
This indicates that traces of the unitary structure are still partially preserved, as the limited number of clusters prevents a uniformly distributed ring. Nevertheless, Fig.~\ref{fig: QFT} \textbf{(c)} shows that this C-D transition marks the threshold beyond which the angular velocity behavior aligns with that of random maps.
More generally, a direct cluster-to-disk transition occurs when the C-R threshold is of the same order as the R-D threshold, $\kappa_{\text{C-R}}\ge\kappa_{\text{R-D}}$. Within the idealized geometric picture underlying Eq.~\eqref{eq: transition solution}, this corresponds to evenly distributed clusters of comparable size with $n \le 4$. The QFT, for which $n=4$ for all system sizes, is therefore the borderline example of this direct-transition scenario.
\par
\textit{Conclusion.---}%
We have shown that the spectrum of integrable systems weakly coupled to an environment exhibits distinct eigenvalue clustering. We demonstrated that dissipation gradually erodes these spectral features that signal the underlying regular dynamics in quantum systems. By analyzing the complex eigenvalue spectra of diluted unitary maps, we identified two key transitions---cluster-to-ring and ring-to-disk---which signal the breakdown of integrable dynamics. 
\par
We introduced the eigenvalue angular velocity as a sensitive diagnostic of the cluster-to-ring transition, providing a quantitative estimate for the critical dissipation threshold. 
These results are universal in the sense that the threshold depends only on the number of clusters and the rank of the dissipative channel and apply across a variety of systems, including Clifford circuits, spin chains, free-fermion circuits, and the quantum Fourier transform.
\par
Finally, in the End Matter we compute the CSRs and show that---because they resolve primarily the intracluster scale---they remain insensitive to the cluster-to-ring transition, which requires information about the global nature of the map.
\par
These results shed light on how dissipation governs the loss of unitary integrability in open quantum systems, offering a concrete proxy for when structured, integrable dynamics give way to universal, chaotic behavior. In typical dynamical systems, the number of spectral clusters---linked to conserved quantities---is expected to grow exponentially with system size, while the number of dissipative channels generally increases only algebraically due to the local nature of most environmental couplings. As a result, in the thermodynamic limit, even exponentially small dissipation is sufficient to erase integrability signatures. 
The finite-system-size analysis reported in the End Matter reinforces this picture as the average number of clusters grows rapidly with system size. In addition, the quantum Fourier transform provides an important complementary example, since its fixed spectral structure places it in the direct cluster-to-disk regime rather than in the generic large-$n$ scaling regime. 
\par
Finally, the strong dependence of the transition on the rank of the dissipative channel suggests that coherence in quantum simulations and computations could be preserved longer by engineering environments with higher-rank noise, thus opening new avenues for mitigating dissipative effects.
\par
\emph{Data availability.---}  The numerical data and simulation code used to generate the figures in this paper are available in the repository of Ref.~\cite{github}.
\par
\emph{Acknowledgments.---}
This work was supported by the Research Council of Norway, project “IKTPLUSS-IKT og digital innovasjon
- 333979” (KW and SD), and by FCT-Portugal, Grant Agreement No. 101017733 (PR, LS, NS, and RP), as part of the QuantERA II project “DQUANT: A Dissipative Quantum Chaos perspective on Near-Term Quantum Computing”. \footnote{\url{https://doi.org/10.54499/QuantERA/0003/2021}} 
PR, NS, and RP acknowledge support from FCT through the financing of the I\&D unit: UID/04540 - Centro de Física e Engenharia de Materiais Avançados. 
RP acknowledges funding from the Deutsche Forschungsgemeinschaft (DFG) under the Priority Programme SPP 2514.
LS was supported by a Research Fellowship from the Royal Commission for the Exhibition of 1851.

\bibliography{combined}

\vspace{-2mm}
\section*{End Matter}

\begin{figure}[tbp!]
    \includegraphics[width=\linewidth]{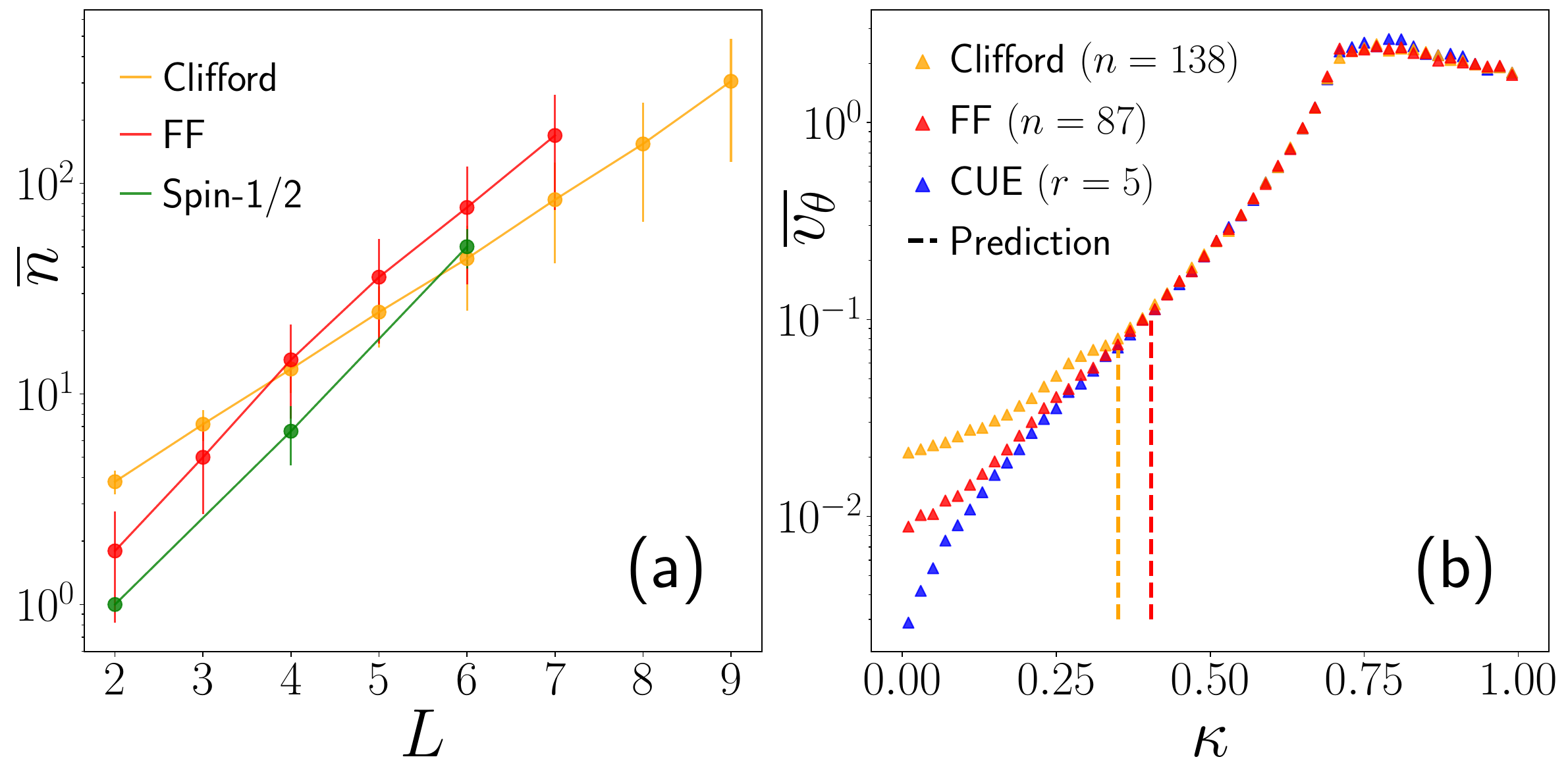}
    \caption{\textbf{(a)}: Average number of clusters as a function of system size. The error bars represent the standard deviation. Each point was averaged over several samples, ranging from $1000$ for lower system sizes to $100$ for higher $L$. \textbf{(b)}: Mean eigenvalue angular velocity for the CUE, free fermions and Clifford models for $L = 7$ and $r = 5$. The dashed lines correspond to the estimate of the transition value $\kappa_{\text{C-R}}$ given by Eq.~\eqref{eq: transition solution}.}
    \label{fig: scaling}
\end{figure}

\emph{Finite-size scaling.---}%
In the main text, we show that the critical value $\kappa_{\text{C-R}}$ depends on the number of spectral clusters. Figure~\ref{fig: scaling} \textbf{(a)} extends this analysis to larger system sizes and directly quantifies the size dependence of the key input $n$. For all three integrable families considered, the average number of clusters increases rapidly with $L$; the data are consistent with an approximately exponential growth. Since Eq.~\eqref{eq: transition solution} depends on $n$ and $r$, this provides the corresponding finite-size trend of the dissipation threshold: as the system size increases, the C-R transition is pushed to progressively smaller values of $\kappa$. The error bars quantify sample-to-sample fluctuations, but they do not alter the overall scaling tendency. The method used to estimate the number of clusters is described in SM~\ref{app: cluster count} and is sufficiently stable to resolve the relevant scaling regime.

\par
Figure~\ref{fig: scaling} \textbf{(b)} complements this analysis by showing the mean eigenvalue angular velocity for $L = 7$. The same mechanism identified at $L = 6$ remains clearly visible at the larger size: the integrable curves approach the CUE curve beyond a finite dissipation scale, and the ordering of the crossover is consistent with the cluster structure of the underlying unitary. For Clifford circuits, where $n$ is known exactly, Eq.~\eqref{eq: transition solution} provides a good quantitative estimate of the crossover. For free-fermion circuits, the predicted value still locates the relevant crossover region and therefore serves as a useful quantitative guide. Taken together, Figs.~\ref{fig: scaling} \textbf{(a)} and \textbf{(b)} show that the transition persists at larger sizes and shifts in the direction anticipated from the dependence on $n$ and $r$.
\par

\emph{Experimental accessibility of spectral dynamics.---}%
The eigenvalue angular velocity introduced above requires tracking individual eigenvalues under small changes of the dissipation strength. This is experimentally demanding, but its prerequisite---the reconstruction of the spectrum of an implemented quantum map---is accessible for few-qubit circuits. We use current quantum hardware to demonstrate that the spectral structures discussed in this work, and their gradual deformation under increasing noise, can be retrieved experimentally.

Quantum map tomography~\cite{wold2024spectra} reconstructs the superoperator implemented by a circuit, e.g., from its responses to Pauli-string inputs. Diagonalizing this reconstructed map gives the complex spectrum of the experimental quantum channel. In the Noisy Intermediate-Scale Quantum (NISQ) regime~\cite{preskillQuantumComputingNISQ2018}, such spectra provide an effective description of the noisy implemented circuit and have been shown to capture non-trivial spectral features~\cite{wold2024spectra,wold2025}.

Here, we implement parametrized three-qubit circuits on the VTT $Q50$ superconducting processor~\cite{vtt}. To increase the effective dissipation, the target circuit was followed by a sequence of $d_{\mathrm{idle}}$ idling blocks. Each idling block consists ideally of a randomly sampled parametrized circuit described by a unitary $U_\ell$ and its inverse $U_\ell^\dagger$, with $\ell=1,\ldots,d_{\mathrm{idle}}$. In the absence of noise these blocks implement the identity; on hardware they add decoherence and gate noise while leaving the ideal unitary action unchanged. Thus $d_{\mathrm{idle}}$ plays the role of an experimentally controlled, monotonic noise parameter: the intrinsic dissipative nature of the NISQ processor is used as a tool to control decoherence. Increasing the number of idling blocks is analogous to increasing $\kappa$ in Eq.~\eqref{eq: five maps}, although extracting a calibrated value of $\kappa$ or an effective Kraus rank is beyond the scope of this work.

Ideally, the target circuit is described by a unitary $U_{\mathrm{circ}}$. If $\Phi_{\mathrm{circ}}\cdot=U_{\mathrm{circ}}\cdot U_{\mathrm{circ}}^\dagger$ denotes the corresponding unitary channel and $\Psi_{\mathrm{idle}}$ denotes the noisy channel generated by the idling sequence, then the reconstructed map can be modeled as $\tilde{\Phi}=\Psi_{\mathrm{idle}}\circ\Phi_{\mathrm{circ}}$. This gives an experimental analogue of the diluted-unitary map in Eq.~\eqref{eq: five maps}, with $U_M=U_{\mathrm{circ}}$ and an effective dissipation strength that increases with $d_{\mathrm{idle}}$.

\begin{figure*}[tbp]
    \includegraphics[width=0.85\linewidth]{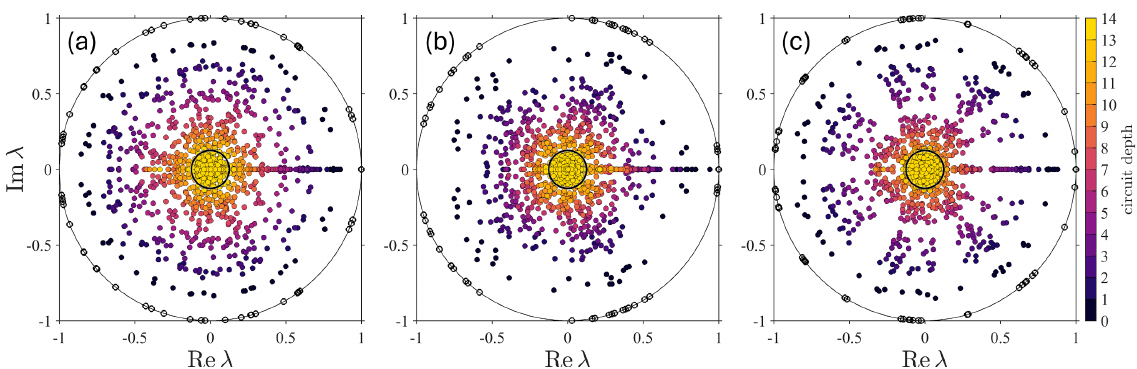}
    \caption{Eigenvalue dynamics retrieved in experiment. Panels \textbf{(a)--(c)} show spectra with distinct clustering features, reconstructed by quantum map tomography~\cite{wold2024spectra} for three three-qubit parametrized target circuits followed by $d_{\mathrm{idle}}$ idling blocks. The number of idling blocks acts as an experimentally controlled monotonic noise parameter. Open circles indicate the spectra of the corresponding ideal target unitaries: \textbf{(a)} a nearly uniformly distributed spectrum on the unit circle and \textbf{(b,~c)} more clustered spectra. As $d_{\mathrm{idle}}$ increases, the reconstructed spectra contract toward the disk expected for a strongly noisy three-qubit map. The inner circle has radius $1/8$, as predicted for full-rank random maps acting on three qubits~\cite{bruzda2009random}. }
    \label{fig:experiment}
\end{figure*}

We selected three representative target circuits---obtained by sampling parametrized random unitary circuits~\cite{wold2024spectra}---whose ideal unitary spectra show distinct clustering features, and performed quantum map tomography for increasing values of $d_{\mathrm{idle}}$. For each depth, a new idling block was added while previously sampled blocks remained fixed. The results, shown in Fig.~\ref{fig:experiment}, reveal that the initial unitary spectral structure (shown as open circles on the unit circle) leaves a visible imprint on the experimental eigenvalue motion before noise washes it out: the circuit with nearly uniformly distributed unitary eigenvalues contracts differently from the circuits with more clustered spectra, while all spectra eventually move toward the disk expected for a strongly noisy three-qubit map, of radius $1/8$, as predicted for full-rank random maps acting on three qubits~\cite{bruzda2009random}. Although the present data are not sufficiently stable to extract reliable angular velocities of individual eigenvalues, the trend that more clustered eigenvalues start interacting with each other for smaller dissipation strength is already visible.

Extracting reliable angular velocities would require reconstructing nearby maps with finer control of the effective noise strength and reduced shot-to-shot fluctuations and device drift. Nevertheless, these experimental results are a proof-of-principle demonstration of spectral-map reconstruction and point to a concrete route toward measuring the angular-velocity diagnostic on more stable few-qubit platforms.

\emph{Complex spacing ratios.---}%
CSRs have proven to be effective indicators of integrable and chaotic non-unitary dynamics. The $n$-th CSR is given by~\cite{sa2020complex} $z_n = (\lambda_n^{\text{NN}} - \lambda_n)/(\lambda_n^{\text{NNN}} - \lambda_n) = r_ne^{i\theta_n}$, where the superscripts NN and NNN denote the nearest and next-to-nearest neighbors of the eigenvalue $\lambda_n$, respectively. For quantum maps drawn from the GinUE, the distribution of $z_n$ in the complex plane shows a characteristic ``bitten-doughnut'' shape. This distribution can be quantified by two key metrics: the average angular component, $-\langle \cos\theta \rangle \approx 0.24$, and the average modulus, $\langle r \rangle \approx 0.74$. In contrast, for systems with uncorrelated eigenvalues, the CSR distribution is uniform in the unit disk, characterized by $-\langle \cos\theta \rangle = 0$ and $\langle r \rangle = 2/3$~\cite{sa2020complex,garcia2022PRX,rubio2022integrability,sa2021integrable,shivam2023many}.

\begin{figure}[tbp!]
    \includegraphics[width=\linewidth]{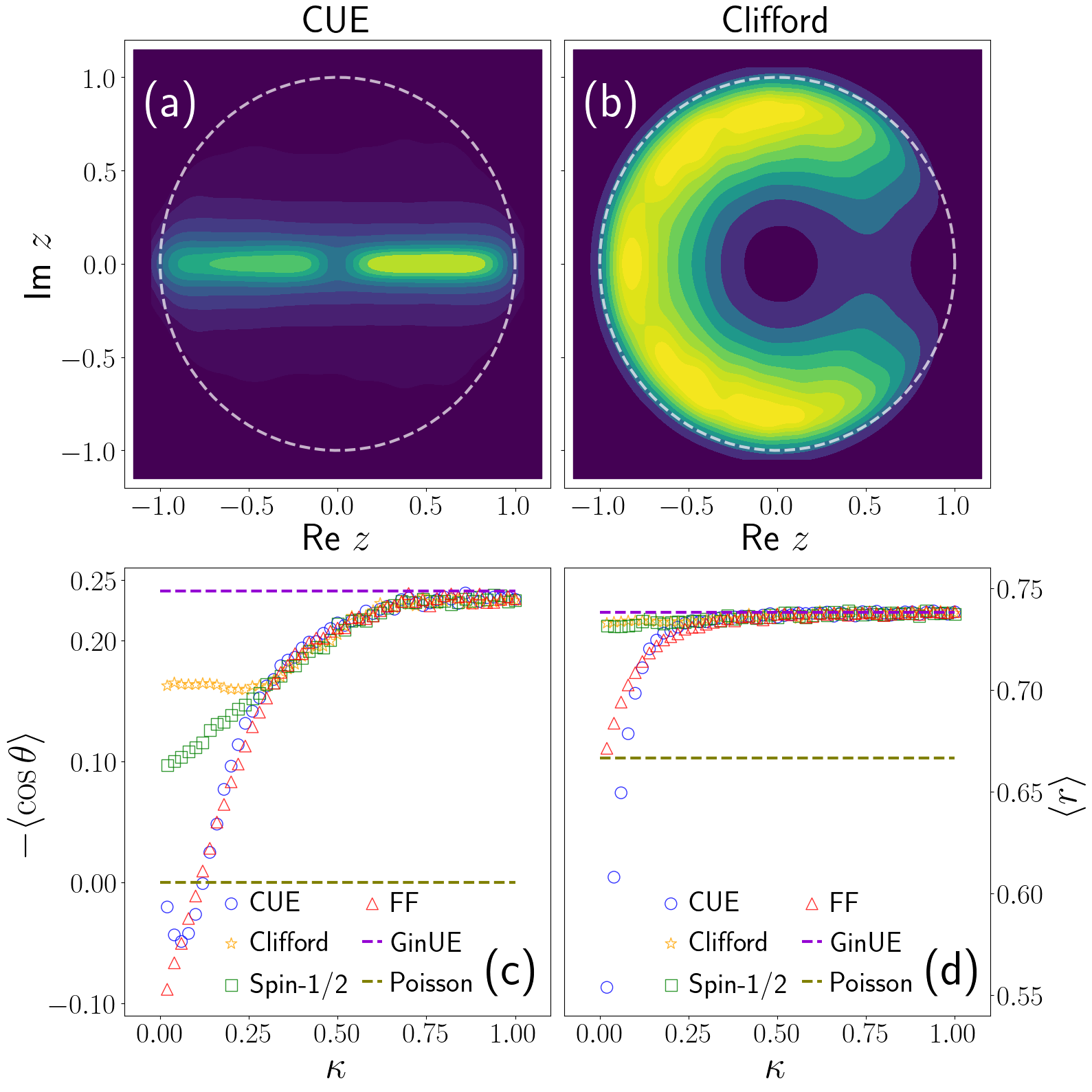}
    \caption{\textbf{(a)} and \textbf{(b)}: Distribution of the CSRs in the complex plane for the CUE \textbf{(a)} and  Clifford \textbf{(b)} models with $\kappa = 0.02$. \textbf{(c)} and \textbf{(d)}: Average $\cos\theta$ \textbf{(c)} and radius \textbf{(d)} of CSRs for all four models described in the main text. We do not present the case of the QFT as it can be considered as a Clifford with four clusters. The dashed lines represent the limits of Poisson and Ginibre statistics. In all the plots, we sampled fifty maps for each model, and used $L = 6$ and $r=5$.}
    \label{fig: CSR}
\end{figure}

Here, we compute the CSRs of the models described in the main text. Figures~\ref{fig: CSR} \textbf{(a)} and \textbf{(b)} respectively show the ratios for the CUE and Clifford models fixing $\kappa = 0.02$. In the case of the CUE, most of the values are distributed around the real axis, as expected---weak dissipation implies that the eigenvalues remain approximately uniformly distributed along the unit circle. In contrast, the Clifford circuit exhibits a distribution consistent with dissipative chaotic behavior. At first glance, this may seem counterintuitive, given that Clifford dynamics are typically regular. 
To understand this, consider the illustrative case of unitary Clifford dynamics in the weakly dissipative regime, $\kappa \ll 1$.
In such systems, the spectrum features highly degenerate eigenvalues. Even a small dissipative term lifts these degeneracies, which can be analyzed using degenerate perturbation theory by diagonalizing the perturbation within the degenerate subspace. Since the perturbation arises from chaotic dissipation, the eigenvalue corrections exhibit GinUE statistics.
CSRs, which depend on both first- and second-nearest-neighbor levels, are particularly sensitive to these corrections.
In contrast, for systems without degeneracies---such as those drawn from the CUE ensemble---dissipation must exceed the mean level spacing to produce significant non-radial corrections. Consequently, CSRs remain largely unaffected at low $\kappa$ in these cases.
This explanation is corroborated by Figures~\ref{fig: CSR} \textbf{(c)} and \textbf{(d)} that show the indicators $\langle r \rangle$ and $-\langle\cos\theta\rangle$ as a function of $\kappa$ for the four models. 
In the $\kappa\to0$ limit, the maps with a larger degree of degeneracy (Clifford and integrable spin-1/2 model) display features consistent with chaotic dynamics at substantially smaller values of $\kappa$ than the less degenerate maps for which these features only arise at larger dissipation strength. As dissipation increases, all models converge toward universal chaotic Ginibre statistics.
\par
In summary, in the main text, we showed that while dissipation breaks the integrability of the underlying unitary map, integrability footprints persist in the spectrum of the resulting dissipative map---even at moderate dissipation strengths. However, these global spectral signatures are not captured by standard spectral measures of chaos, such as the CSRs, which reflect local chaotic statistics within clusters but fail to detect the global cluster-to-ring transition.

%
%

\clearpage
\newpage
\onecolumngrid
\appendix

\setcounter{figure}{0}
\renewcommand{\thefigure}{SM\arabic{figure}}%
\setcounter{page}{1}
\renewcommand{\thepage}{SM-\arabic{page}}%

\begin{center}\Large{
		\textbf{Supplemental Material for}\\
		\textbf{Dissipation-Induced Threshold on Integrability Footprints}\vspace{2ex}
        }
\end{center}

\section{Sampling methods}\label{app: methods}

In this section, we explain the numerical methods used to generate the random matrices described in the main text.

\subsection{Circular unitary ensemble}

 The algorithm to generate a random matrix from the CUE is as follows~\cite{mezzadri2006generate}: we sample a $d\times d$ random matrix $Z$ from the GinUE, and perform any QR decomposition algorithm such that $Z = QR$. Then, write the diagonal matrix $\Lambda = \text{diag} (R_{11}/|R_{11}|, \dots,R_{dd}/|R_{dd}|)$, where $R_{ii}$ is the $i$-th diagonal element of $R$. Finally, we realize the transformation $Q \to Q\Lambda$. The new $Q$ is Haar-distributed and therefore belongs to the CUE.

\subsection{Kraus operators}

Let us now look at how to sample the $r$ Kraus operators $K_j$. A way to proceed is by following Ref.~\cite{bruzda2009random}. To that end, we sample a matrix $W \in \text{CUE}$ with dimensions $rd \times rd$. Then, the Kraus operators correspond to the first (or any) block column of $d \times d$ matrices. The condition $\sum_{j=1}^r K_j^{\dagger}K_j = \mathbb{I}$ is satisfied due to the orthogonality of the columns of $W$. However, this algorithm is computationally inefficient since it only uses the information of one block column of an enlarged matrix. Thus, in the numerics of the paper, we alternatively considered the following algorithm.
\par
We begin by generating an $rd \times d$ random matrix from the GinUE, perform a $QR$ decomposition, and extract the rectangular matrix $Q$, which is also $rd \times d$. Then we perform the transformation $Q \to Q\Lambda$, with $\Lambda = \text{diag} (R_{11}/|R_{11}|,\dots,R_{dd}/|R_{dd}|)$, necessary to obtain the correct eigenvalue density distribution. Finally, we take $d$ consecutive rows and truncate the matrix dividing it into $r$ $d \times d$ matrices:
\begin{equation}
  Q = 
  \left( 
      \begin{array}{ccc}
        \begin{array}{ccc}
          Q_{1,1} & \cdots & Q_{1,d}
          \\ 
          \vdots  & \ddots & \vdots
          \\
          Q_{d,1} & \cdots & Q_{d,d}
          \\
          \hline
          Q_{d+1,1} & \cdots & Q_{d+1,d}
          \\ 
          \vdots  & \ddots & \vdots
          \\
          Q_{2d,1} & \cdots & Q_{2d,d}
           \\
          \hline
          \vdots  & \ddots & \vdots
          \\
          Q_{rd,1} & \cdots & Q_{rd,d}
        \end{array} 
      \end{array}
    \right)
  ~ 
  \begin{array}{c}
    \left.\right.
    \\
    \left.\kern-\nulldelimiterspace
    \vphantom{
    \begin{array}{ccc}
          Q_{1,1} & \cdots & Q_{1,d}
          \\ 
          \vdots  & \ddots & \vdots
          \\
          Q_{d,1} & \cdots & Q_{d,d}
    \end{array}
    }
    \right\} K_1 \\
    \left.\kern-\nulldelimiterspace
    \vphantom{
    \begin{array}{ccc}
          Q_{d+1,1} & \cdots & Q_{d+1,d}
          \\ 
          \vdots  & \ddots & \vdots
          \\
          Q_{2d,1} & \cdots & Q_{2d,d}
    \end{array} 
    }
    \right\} K_2
    \\
    \left.\kern-\nulldelimiterspace
    \vphantom{
    \begin{array}{ccc}
          Q_{d+1,1} & \cdots & Q_{d+1,d}
          \\ 
          \vdots  & \ddots & \vdots
          \\
          Q_{2d,1} & \cdots & Q_{2d,d}
        \end{array} 
    }
    \right. \vdots
  \end{array}.
\end{equation}
The condition $\sum_{j=1}^r K_j^{\dagger}K_j = \mathbb{I}$ is automatically met since $Q^{\dagger}Q = \mathbb{I}$. More explicitly:
\begin{equation}
    \left(Q^{\dagger}Q\right)_{\alpha \alpha'} 
    = \sum_{(\beta \gamma)=1}^{rd} \big(Q^{\dagger}\big)_{\alpha (\beta \gamma)} Q_{(\beta \gamma) \alpha'} 
    = \sum_{\beta=1}^r \sum_{\gamma=1}^d \big(K^{\dagger}_{\beta}\big)_{\alpha \gamma} \big(K_{\beta}\big)_{\gamma \alpha'} 
    = \sum_{\beta=1}^r \left(K^{\dagger}_{\beta} K_{\beta}\right)_{\alpha \alpha'}
    =\left(\sum_{\beta=1}^r K^{\dagger}_{\beta} 
    K_{\beta}\right)_{\alpha \alpha'}
    = \mathbb{I}_{\alpha \alpha'}.
\end{equation}

 \subsection{Clifford operator}

 We uniformly sample a Clifford gate according to Ref.~\cite{van2021simple}, which can be done efficiently, i.e., in polynomial time. The algorithm takes advantage of the tableau representation of a Clifford operator $\mathcal{C}$~\cite{aaronson2004improved}, where acting with Clifford operations translates into actions on the tableau, e.g., a Hadamard gate is responsible for exchanging columns. 
\par
It is worth noting that, since the presence of dissipation does not respect the Clifford structure we have to consider a generic map, which requires explicitly computing a $2^L\times2^L$ matrix. Thus, even though the sampling can be done in polynomial time, the computation of the dissipative map scales exponentially with system size and is no longer computationally efficient.

\subsection{Integrable spin-1/2 chain}

 We will consider the work of Ref.~\cite{vanicat2018integrable}, which introduces a spin-1/2 chain that is integrable in the Bethe-ansatz sense~\cite{XXX_model}. 
 This model corresponds to taking a brickwall circuit with periodic boundary conditions, where each two-qubit unitary $u_{j,j+1}$ acting on qubits $j$ and $j+1$ is taken to be the $\check{{R}}$ matrix\footnote{This matrix corresponds to a solution of the Yang-Baxter equation~\cite{baxter1973asymptotically, jimbo1985q}.} of the XXX model~\cite{XXX_model},
\begin{equation}
    u_{j, j+1} = \check{R}_{j, j+1}(\delta),
    \quad\quad
        \check{R}(\delta) = \frac{\mathbb{I} + i\delta P}{\mathbb{I} + i\delta},
\end{equation}
where the real parameter $\delta$ is a spin coupling constant and $P_{j,j+1} = 1/2(\mathbb{I} + \boldsymbol{\sigma}_j\cdot\boldsymbol{\sigma}_{j+1})$, with $\boldsymbol{\sigma} = (\sigma^x, \sigma^y, \sigma^z)$ a vector with the Pauli matrices. A one-layer unitary matrix $\mathcal{U}$ acting on $L$ qubits is obtained by a two-step operation, first consisting of the action of $u$ over the odd-even sites, and then the action over the even-odd sites,
\begin{equation}
    \mathcal{U} = \mathcal{U}_{\text{even}}\mathcal{U}_{\text{odd}} = \prod_{j=1}^{L/2} u_{2j-1, 2j} \prod_{k=1}^{L/2} u_{2k, 2k+1},
\end{equation}
where we must restrict to even $L$ and the sites $1$ and $L+1$ are the same, as imposed by periodic boundary conditions. Let us call $\mathcal{U}_t$ the operation realized in the layer $t$. The total unitary $U$ corresponds to multiplying the matrices over $T$ layers:
\begin{equation}
    U = \prod_{t=1}^T \mathcal{U}_t.
\end{equation}
To randomly sample a unitary matrix $\mathcal{U}_t$, we obtain a random $\theta \in [0, \pi]$ and calculate $\delta = \tan\theta$, allowing us to compute a random real number, even though not uniformly. This parameterization satisfies the relation $\check{R}\left(\tan\theta\right) \check{R}\left(\tan\theta'\right) = \check{R}\left[\tan(\theta+\theta')\right]$. 
\par
For numerical simulations, we fixed the number of layers in $T = L$ and considered the case where each layer consists of the same operation, i.e. $\mathcal{U}_t \equiv \mathcal{U}$, such that the total matrix corresponds to $U = \mathcal{U}^L$. With this method, we only sample $\theta$ once.

\subsection{Free fermions}

The most general quadratic fermionic Hamiltonian can be written as
\begin{equation}\label{eq: general quadratic hamiltonian}
    H_{FF} = \sum_{j, k=1}^L 2h_{jk}c_j^{\dagger}c_k + \sum_{j, k=1}^L \left( \Delta_{jk}c_j^{\dagger}c_k^{\dagger} - \Delta_{jk}^* c_j c_k \right),
\end{equation}
where $L$ is the number of sites. The first term describes the dynamics of the fermions within the chain, and the last two refer to the creation and annihilation of pairs, which causes the non-conservation of the number of particles. However, parity is conserved, i.e., whether the number of particles is even or odd never changes.
\par
Defining the Nambu vector $C = (c_1 \; \dots \; c_L \: c_1^{\dagger} \; \dots \; c_L^{\dagger})^T$ we can write Eq.~\eqref{eq: general quadratic hamiltonian} as
\begin{equation}\label{eq: compact quadratic hamiltonian}
    H_{FF} = C^{\dagger} \boldsymbol{H} C\text{, with  } 
    \boldsymbol{H} = 
    \begin{pmatrix}
        h & \Delta
        \\
        -\Delta^{*} & -h^*
    \end{pmatrix},
\end{equation}
where $\boldsymbol{H}$ is the single particle Hamiltonian, and $h = h^{\dagger}$ and $\Delta = -\Delta^T$ are $L\times L$ matrices.
\par
We can map fermionic operators onto spin operators through the Jordan-Wigner transformation~\cite{somma2002simulating, nielsen2019fermionic}
\begin{equation}
    \begin{split}
        &c^\dagger_m =e^{i\frac{\pi}{2}\sum_{k=1}^{j-1}(1-\sigma_k^z)} \sigma^-_m = S_{1m}\sigma^-_m, \\
        &c_m = e^{-i\frac{\pi}{2}\sum_{k=1}^{m-1}(1-\sigma_k^z)} \sigma^+_m=S^\dagger_{1m}\sigma^+_m,
    \end{split}
\end{equation}
which reproduces the correct commutation relations with
\begin{equation}
    c_m^{\dagger}c_m = \frac{1}{2} (1 - \sigma_m^z).
\end{equation}
    With this definition, the string operator is $S_{ij}=S^\dagger_{ij}=(-1)^{p_{i,j}}$, where $p_{i,j}$ counts the number of spins $\ket{-}$ between the $i$-th and $(j-1)$-th positions. Defining the spinor
\begin{equation}
    \mathbf{\Sigma}=\begin{pmatrix}\sigma^-_{1}\\\vdots\\\sigma^-_{N}\\\sigma^+_{1}\\\vdots\\\sigma^+_{N}\end{pmatrix}=\begin{pmatrix}\mathbf{\sigma}^-\\\mathbf{\sigma}^+\end{pmatrix},
\end{equation}
we can encode the Jordan-Wigner transformation into a $2L\times 2L$ matrix and write the Nambu vector as
\begin{equation}\label{eq: JW matrix}
    C = J\mathbf{\Sigma},\quad 
    J=
    \begin{pmatrix}
        \mathbf{0} & \mathbf{S}
        \\
        \mathbf{S} & \mathbf{0} 
    \end{pmatrix},
 \quad \mathbf{S}=\text{diag}\{S_{11},\dots,S_{1N}\},
\end{equation}
where $S_{ij}$ is the string operator. If we plug Eq.~\eqref{eq: JW matrix} in Eq.~\eqref{eq: compact quadratic hamiltonian} we obtain the spin first-quantized Hamiltonian
\begin{equation}
    H_{FF}=\mathbf{\Sigma}^\dagger J\boldsymbol{H}J \mathbf{\Sigma} = \mathbf{\Sigma}^\dagger \boldsymbol{H}_{\text{spin}} \mathbf{\Sigma}, \quad \boldsymbol{H}_{\text{spin}}= 
    \begin{pmatrix}
  -\mathbf{S}\mathbf{h}^*\mathbf{S} & -\mathbf{S}\mathbf{\Delta}^*\mathbf{S} \\ 
  \mathbf{S}\mathbf{\Delta}\mathbf{S} & \mathbf{S}\mathbf{h}\mathbf{S}
 \end{pmatrix},
\end{equation}
with
\begin{equation}
    \mathbf{S}\mathbf{h}\mathbf{S}=\mathbf{h}_{\text{spin}}
    =
    \begin{pmatrix}
h_{11} & h_{12}(-1)^{p_{12}} & \hdots & h_{12}(-1)^{p_{1N}} \\
h^*_{12}(-1)^{p_{12}} & h_{22} & \hdots & \vdots \\
\vdots & \vdots & \ddots & h_{N-1N}(-1)^{p_{NN-1}} \\
h^*_{12}(-1)^{p_{1N}} & \hdots &  h^*_{N-1N}(-1)^{p_{NN-1}} & h_{NN}\ , 
\end{pmatrix},
\end{equation}
and
\begin{equation}
\mathbf{S}\mathbf{\Delta}\mathbf{S}=\mathbf{\Delta}_{\text{spin}}=
    \begin{pmatrix}
0 & \Delta_{12}(-1)^{p_{12}} & \hdots & \Delta_{12}(-1)^{p_{1N}} \\
-\Delta_{12}(-1)^{p_{12}} & 0 & \hdots & \vdots \\
\vdots & \vdots & \ddots & \Delta_{N-1N}(-1)^{p_{NN-1}} \\
-\Delta_{12}(-1)^{p_{1N}} & \hdots & -\Delta_{N-1N}(-1)^{p_{NN-1}} & 0 
\end{pmatrix}.   
\end{equation}

We proceed by computing the matrix elements of the many-body Hamiltonian in the many-body basis:

\begin{eqnarray}
    \boldsymbol{\sigma}^-\mathbf{h}_{\text{spin}}\boldsymbol{\sigma}^+\ket{\sigma_1,\dots,\sigma_k,\dots,\sigma_l,\dots\sigma_N}&=&\Bigl(\sum_{k=1}^N h_{kk}\sigma^-_k\sigma^+_k + \sum_{k<l}h_{kl}\sigma^-_{k}(-1)^{p_{kl}}\sigma^+_l\nonumber \\
    &+& \sum_{k<l}h^*_{kl}\sigma^-_l(-1)^{p_{kl}}\sigma^+_k\Bigr)\ket{\sigma_1,\dots,\sigma_k,\dots,\sigma_l,\dots\sigma_N}\nonumber\\
    &=&\Bigl(\sum_{k=1}^N h_{kk}\delta_{\sigma_k,-}\Bigr)\ket{\sigma_1,\dots,\sigma_k,\dots,\sigma_l,\dots\sigma_N}\\
    &+&\sum_{k<l}h_{kl}\delta_{\sigma_k,+}\delta_{\sigma_l,-}(-1)^{p_{k+1,l}}\ket{\sigma_1,\dots,-,\dots,+,\dots\sigma_N}\nonumber\\
    &+&\sum_{k<l}h^*_{kl}\delta_{\sigma_k,-}\delta_{\sigma_l,+}(-1)^{p_{k+1,l}}\ket{\sigma_1,\dots,+,\dots,-,\dots\sigma_N},\nonumber
\end{eqnarray}
\begin{eqnarray}
    -\boldsymbol{\sigma}^+\mathbf{h}^*_{\text{spin}}\boldsymbol{\sigma}^-\ket{\sigma_1,\dots,\sigma_k,\dots,\sigma_l,\dots\sigma_N}&=&\Bigl(-\sum_{k=1}^N h_{kk}\sigma^+_k\sigma^-_k - \sum_{k<l}h^*_{kl}\sigma^+_{k}(-1)^{p_{kl}}\sigma^-_l\nonumber \\
    &-& \sum_{k<l}h_{kl}\sigma^+_l(-1)^{p_{kl}}\sigma^-_k\Bigr)\ket{\sigma_1,\dots,\sigma_k,\dots,\sigma_l,\dots\sigma_N}\nonumber\\
    &=&-\Bigl(\sum_{k=1}^N h_{kk}\delta_{\sigma_k,+}\Bigr)\ket{\sigma_1,\dots,\sigma_k,\dots,\sigma_l,\dots\sigma_N}\\
    &+&\sum_{k<l}h^*_{kl}\delta_{\sigma_k,-}\delta_{\sigma_l,+}(-1)^{p_{k+1,l}}\ket{\sigma_1,\dots,+,\dots,-,\dots\sigma_N}\nonumber\\
    &+&\sum_{k<l}h_{kl}\delta_{\sigma_k,+}\delta_{\sigma_l,-}(-1)^{p_{k+1,l}}\ket{\sigma_1,\dots,-,\dots,+,\dots\sigma_N},\nonumber
\end{eqnarray}
\begin{eqnarray}
    \boldsymbol{\sigma}^-\mathbf{\Delta}_{\text{spin}}\boldsymbol{\sigma}^-\ket{\sigma_1,\dots,\sigma_k,\dots,\sigma_l,\dots\sigma_N}&=&\sum_{k<l}\Delta_{kl}\sigma^-_{k}(-1)^{p_{kl}}\sigma^-_l
    -\Delta_{kl}\sigma^-_l(-1)^{p_{kl}}\sigma^-_k\Bigr)\ket{\sigma_1,\dots,\sigma_k,\dots,\sigma_l,\dots\sigma_N}\nonumber\\
    &=& \sum_{k<l}2\Delta_{kl}\delta_{\sigma_k,+}\delta_{\sigma_l,+}(-1)^{p_{k+1,l}}\ket{\sigma_1,\dots,-,\dots,-,\dots\sigma_N},\nonumber\\
\end{eqnarray}
\begin{eqnarray}
    -\boldsymbol{\sigma}^+\mathbf{\Delta}^*_{\text{spin}}\boldsymbol{\sigma}^+\ket{\sigma_1,\dots,\sigma_k,\dots,\sigma_l,\dots\sigma_N}&=&-\sum_{k<l}\Delta^*_{kl}\sigma^+_{k}(-1)^{p_{kl}}\sigma^+_l
    +\Delta^*_{kl}\sigma^+_l(-1)^{p_{kl}}\sigma^+_k\Bigr)\ket{\sigma_1,\dots,\sigma_k,\dots,\sigma_l,\dots\sigma_N}\nonumber\\
    &=&  \sum_{k<l}2\Delta^*_{kl}\delta_{\sigma_k,-}\delta_{\sigma_l,-}(-1)^{p_{k+1,l}}\ket{\sigma_1,\dots,+,\dots,+,\dots\sigma_N}.\nonumber\\
\end{eqnarray}
Joining all these results finally gives the matrix elements
\begin{equation}\label{eq: appendix H_FF}
    \begin{split}
&\mel{\sigma_1'\dots,\sigma_k',\dots,\sigma_l',\dots,\sigma_N'}{H_{FF}}{\sigma_1\dots,\sigma_k,\dots,\sigma_l,\dots,\sigma_N}\\[10pt]
=&
\Bigl[\Bigl(\sum_{k=1}^N h_{kk}(\delta_{\sigma_k,-}-\delta_{\sigma_k,+})\Bigr)\delta_{\sigma_k',\sigma_k}\delta_{\sigma_l',\sigma_l}\\
+&
\sum_{k<l}2(-1)^{p_{k+1,l}}\Bigl((h_{kl}\delta_{\sigma_k,+}\delta_{\sigma_l,-}\delta_{\sigma_k',-}\delta_{\sigma_l',+}+h^*_{kl}\delta_{\sigma_k,-}\delta_{\sigma_l,+}\delta_{\sigma_k',+}\delta_{\sigma_l',-}
\\
+&
\Delta_{kl}\delta_{\sigma_k,+}\delta_{\sigma_l,+}\delta_{\sigma_k',-}\delta_{\sigma_l',-}+\Delta_{kl}\delta_{\sigma_k,-}\delta_{\sigma_l,-}\delta_{\sigma_k',+}\delta_{\sigma_l',+}\Bigr)\Bigr]\prod_{m\neq k,l}\delta_{\sigma_m',\sigma_m}.
    \end{split}
\end{equation}
The unitary matrix is obtained by
\begin{equation}
    U_{FF} = e^{-i H_{FF}},
\end{equation}
with $H_{FF}$ computed through Eq.~\eqref{eq: appendix H_FF}.
\par
The random quadratic Hamiltonian is obtained as follows. We can write Eq.~\eqref{eq: compact quadratic hamiltonian} in terms of Majorana fermions
\begin{equation}
    \gamma = VC, \quad V = \frac{1}{\sqrt{2}}
    \begin{pmatrix}
        \mathbb{I} & \mathbb{I}
        \\
        -i\mathbb{I} & i\mathbb{I}
    \end{pmatrix}.
\end{equation}
In this basis, the first quantized Hamiltonian $\boldsymbol{H}_{\text{Majo}}=V\boldsymbol{H}V^{\dagger}$ is a purely imaginary antisymmetric matrix. Therefore, $u=e^{i\boldsymbol{H}_{\text{Majo}}}$ belongs to the special orthogonal (SO) group. Hence:
\begin{enumerate}
    \item Generate $u\in$ SO($2L$).
    \item Obtain $\boldsymbol{H}_{\text{Majo}}=i\log u$.
    \item Transform it to complex fermions $\boldsymbol{H}=V^{\dagger}\boldsymbol{H}_{\text{Majo}}V$, and identify $h$ and $\Delta.$
    \item Use Eq. \eqref{eq: appendix H_FF}.
\end{enumerate}

\section{Fidelity cannot detect the loss of integrability}\label{app: fidelity}

In this section, we show that the average fidelity is not sensitive to the regular or chaotic nature of the faultless unitary evolution. To do so, we will first write an expression of the fidelity, then we shall rewrite it in a convenient form to prove that the left-right invariance of the Haar measure is behind this insensitivity.

The outcome state of the faultless circuit is
\begin{equation}
    \ket{\Psi}=U_TU_{T-1}\cdots U_1\ket{\psi_0},
\end{equation}
where $U_\tau$  is the unitary layer at time $\tau$, that can be structured or unstructured, fixed or sampled independently. 
The dissipative circuit produces a mixed state
\begin{equation}
\rho=\Phi(\Phi(\dots\Phi(\ket{\psi_0}\bra{\psi_0})))=\sum_{\mu_1,\dots,\mu_T=0}^{r}\overleftarrow{\prod_{\tau=1}^T}\hat{K}_{\mu_\tau}\ket{\psi_0}\bra{\psi_0}\overrightarrow{\prod_{\tau=1}^T}\hat{K}^\dagger_{\mu_\tau},
\end{equation}
where we implement $T$ times the quantum channel $\Phi(\rho_\tau)$ which is given by the \emph{diluted unitary}
\begin{equation}
    \Phi(\rho_{\tau+1})=\sum_{\mu_{\tau+1}=0}^r\hat{K}_{\mu_{\tau+1}}\rho_\tau\hat{K}^\dagger_{\mu_{\tau+1}},
\end{equation}
where 
\begin{align}
    &\hat{K}_0=\sqrt{1-\kappa}\ U\ , \quad \text{ verifying } \quad U^\dagger\tilde{U}=\mathds{1},\\
    &\hat{K}_{\mu\neq0}=\sqrt{\kappa}\ K_\mu\ ,\quad \text{ verifying } \quad \sum_{\mu=1}^{r}K^\dagger_\mu, K_\mu=\mathds{1}
\end{align}
for $\kappa\in[0,1]$. We are interested in computing the average of the fidelity
\begin{equation}
    \mathcal{F}=\mel{\Psi}{\rho}{\Psi}.
    \label{eq:sm_fid}
\end{equation}

To do so, it is useful to use the vectorized notation:
\begin{equation}
  \kket{\psi\phi} =\ket{\psi}\otimes\ket{\phi^{T}}
\end{equation}

Equipped with it, we can write the fidelity in the 4-copy Hilbert space $\mathcal{H}(2^{4L})$: 
\begin{align}
\mathcal{F}= & \mel{\Psi}{\rho}{\Psi}\nonumber\\
= & \bra{\psi_{0}}\left(\overrightarrow{\prod_{\tau=1}}U_{\tau}^{\dagger}\right)\left[\sum_{\mu_1,\dots,\mu_T=0}^{r}\left(\overleftarrow{\prod_{\tau=1}}\hat{K}_{\mu_\tau}\right)\ket{\psi_0}\bra{\psi_0}\left(\overrightarrow{\prod_{\tau=1}}\hat{K}^\dagger_{\mu_\tau}\right)\right]\left(\overleftarrow{\prod_{\tau=1}}U_{\tau}\right)\ket{\psi_0}\nonumber \\
= & \sum_{\bs{\mu}}\sum_{\bs{nm}}\bra{\psi_{0}}\left(\overrightarrow{\prod_{\tau=1}}U_{\tau}^{\dagger}\right)\ket{\bs m}\bra{\bs m}\left(\overleftarrow{\prod_{\tau=1}}\hat{K}_{\mu_\tau}\right)\ket{\psi_0}\bra{\psi_0}\left(\overrightarrow{\prod_{\tau=1}}\hat{K}^\dagger_{\mu_\tau}\right)\ket{\bs n}\bra{\bs n}\left(\overleftarrow{\prod_{\tau=1}}U_{\tau}\right)\ket{\psi_{0}}\nonumber\\
= & \sum_{\bs{\mu}}\sum_{\bs{nm}}\bra{\psi_{0}\bs{m}\psi_0\bs{n}}\left(\overrightarrow{\prod_{\tau=1}}U_{\tau}^{\dagger}\right)\otimes\left(\overleftarrow{\prod_{\tau=1}}\hat{K}_{\mu_\tau}\right)\otimes\left(\overrightarrow{\prod_{\tau=1}}\hat{K}^\dagger_{\mu_\tau}\right)\otimes\left(\overleftarrow{\prod_{\tau=1}}U_{\tau}\right)\ket{\bs{m}\psi_{0}\bs{n}\psi_0}\nonumber\\
= & \sum_{\bs{\mu}}\sum_{\bs{nm}}\bbra{\bs{m}\bs{n}\bs{n}\bs{m}}\left(\overleftarrow{\prod_{\tau=1}}\hat{K}_{\mu_\tau}\right)\otimes\left(\overrightarrow{\prod_{\tau=1}}\hat{K}^\dagger_{\mu_\tau}\right)^T\otimes\left(\overleftarrow{\prod_{\tau=1}}U_{\tau}\right)\otimes\left(\overrightarrow{\prod_{\tau=1}}U_{\tau}^{\dagger}\right)^T\kket{\psi_{0}\psi_{0}\psi_0\psi_{0}}\nonumber\\
= & \sum_{\bs{\mu}}\sum_{\bs{nm}}\bbra{\bs{m}\bs{n}\bs{n}\bs{m}}\overleftarrow{\prod_{\tau=1}}\left( \hat{K}_{\mu_\tau}\otimes\hat{K}^*_{\mu_\tau}\otimes U_{\tau}\otimes U_{\tau}^{*}\right)\kket{\psi_{0}\psi_{0}\psi_0\psi_{0}}\nonumber \\
= &\frac{1}{d}\sum_{\bs{lnm}}\bbra{\bs{m}\bs{n}\bs{n}\bs{m}}\overleftarrow{\prod_{\tau=1}}\Bigl((1-\kappa)U_\tau\otimes U^*_{\tau}\otimes U_{\tau}\otimes U_{\tau}^{*}+\kappa\sum_{\mu_\tau=1}^{r}K_{\mu_\tau}\otimes K^*_{\mu_\tau}\otimes U_{\tau}\otimes U_{\tau}^{*} \Bigr)\kket{\bs{l}\bs{l}\bs{l}\bs{l}},
\end{align}
where 
\begin{equation}
    \sum_{\bs n}\ket{\bs{n}}=\left(\sum_{m=0,1}\ket{m}\right)^{\otimes L},
\end{equation}
and $\bs{\mu}=(\mu_T,\dots, \mu_1)$ with $\mu_\tau=0,\dots,r$. Observe that in the last equality, we have further averaged over the computational basis
$$
\kket{\psi_0\psi_0\psi_0\psi_0}\ \rightarrow\ \frac{1}{d}\sum_{\bm{l}}\kket{\bm{l}\bm{l}\bm{l}\bm{l}}.
$$
\begin{equation}
    \mathcal{F}=\frac{1}{d}\bbra{\pone}\overleftarrow{\prod_{\tau=1}}\Bigl((1-\kappa)\bu_\tau+\kappa\bk_\tau\Bigr)\kket{\pthree},\quad \begin{cases}
    \kket{\pone}& =\sum_{\bs{nm}}\kket{\bs{m}\bs{n}\bs{n}\bs{m}}\\
    \bu_\tau&= U_\tau\otimes U^*_\tau\otimes U_\tau\otimes U^*_\tau\\
    \bk_\tau &=\sum_{\mu_\tau=1}^{r}K_{\mu_\tau}\otimes K^*_{\mu_\tau}\otimes U_\tau\otimes U*_\tau\\
    \kket{\pthree}& =\sum_{\bs{l}}\kket{\bs{l}\bs{l}\bs{l}\bs{l}}
    \end{cases}
\end{equation}
Then, the fidelity can be written as:
\begin{equation}
\label{eq:F_SM_toprove}
\begin{split}
    \mathcal{F} &=\frac{1}{d} \bbra{\pone}\overleftarrow{\prod_{\tau=1}}\bu_\tau\overleftarrow{\prod_{\tau=1}}\Bigl((1-\kappa)\mathds{1}+\kappa \overrightarrow{\prod^{\tau}_{\tau'=1}}\bu^\dagger_{\tau'}\bk_\tau\overleftarrow{\prod^{\tau-1}_{\tau''=1}}\bu_{\tau''}\Bigr)\kket{\pthree}\\
    &=\frac{1}{d}\bbra{\pone}\bu(T,1)\prod^{T-1}_{\tau=0}\Bigl((1-\kappa)\mathds{1}+\kappa \bu^\dagger(T-\tau,1)\bk_{T-\tau}\bu(T-1-\tau,1)\Bigr)\kket{\pthree},
    \end{split}
\end{equation}
where we used 
\begin{equation}
    \overrightarrow{\prod^{\tau}_{\tau'}}\bu_{\tau'}\equiv \bu(\tau,\tau').
\end{equation}

Equation~(\ref{eq:F_SM_toprove}) can be easily proved by induction. Consider the case $T=2$:
\begin{align}
    \mathcal{F}&=\frac{1}{d}\bbra{\pone}\Bigl((1-\kappa)\bu_2+\kappa\bk_2\Bigr)\Bigl((1-\kappa)\bu_1+\kappa\bk_1\Bigr)\kket{\pthree}\nonumber\\
    &=\frac{1}{d}\bbra{\pone}\bu_2\Bigl((1-\kappa)\bu_1+\kappa\bu^\dagger_2\bk_2\bu_1\Bigr)\Bigl((1-\kappa)+\kappa\bu^\dagger_1\bk_1\Bigr)\kket{\pthree}\nonumber\\
    &=\frac{1}{d}\bbra{\pone}\bu_2\bu_1\Bigl((1-\kappa)\mathds{1}+\kappa\bu^\dagger_1\bu^\dagger_2\bk_2\bu_1\Bigr)\Bigl((1-\kappa)\mathds{1}+\kappa\bu^\dagger_1\bk_1\Bigr)\kket{\pthree}.
\end{align}
Then, the case $T-1$ is
\begin{equation}
    \mathcal{F}=\frac{1}{d}\bbra{\pone}\bu(T-1,1)\prod^{T-2}_{\tau=0}\Bigl((1-\kappa)\mathds{1}+\kappa \bu^\dagger(T-1-\tau,1)\bk_{T-1-\tau}\bu(T-2-\tau,1)\Bigr)\kket{\pthree}
\end{equation}
and the case $T$ is
\begin{align}
    \mathcal{F}&=\frac{1}{d}\bbra{\pone}\Bigl((1-\kappa)\bu_T+\kappa\bk_T\Bigr)\bu(T-1,1)\prod^{T-2}_{\tau=0}\Bigl((1-\kappa)\mathds{1}+\kappa \bu^\dagger(T-1-\tau,1)\bk_{T-1-\tau}\bu(T-2-\tau,1)\Bigr)\kket{\pthree}\nonumber\\
    &=\frac{1}{d}\bbra{\pone}\bu_T\bu(T-1,1)\Bigl((1-\kappa)\mathds{1}+\kappa \bu^\dagger(T-1,1)\bk_T\bu(T-1,1)\Bigr)\nonumber\\
    &\hspace{5cm}\times \prod^{T-2}_{\tau=0}\Bigl((1-\kappa)\mathds{1}+\kappa \bu^\dagger(T-1-\tau,1)\bk_{T-1-\tau}\bu(T-2-\tau,1)\Bigr)\kket{\pthree}.
\end{align}

Equation~(\ref{eq:F_SM_toprove}) is quite advantageous since
\begin{equation}
    \bbra{\pone}\bu(T,1)=\bbra{\pone}.
\end{equation}
To see this, we undo the vectorization:
\begin{align}
\bbra{\pone}\bu(T,1)&=\sum_{\bs{nm}}\bbra{\bs{m}\bs{n}\bs{n}\bs{m}}\Bigl( U(T,1)\otimes U^*(T,1)\otimes U(T,1)\otimes U^*(T,1)\Bigr)\nonumber \\
&=\sum_{\bs{nm}}\bbra{\bs{m}\bs{m}\bs{n}\bs{n}}\Bigl( U(T,1)\otimes U^*(T,1)\otimes U(T,1)\otimes U^*(T,1)\Bigr)\nonumber \\ 
&=\sum_{\bs{nm}}\Bigl( U^\dagger(T,1)\otimes U^T(T,1)\otimes U(T,1)^\dagger\otimes U^T(T,1)\kket{\bs{m}\bs{m}\bs{n}\bs{n}}\Bigr)^\dagger\nonumber \\ 
&=\sum_{\bs{nm}}\Bigl( U^\dagger(T,1)\ket{\bs{m}}\bra{\bs{m}}U(T,1)\otimes U^\dagger(T,1)\ket{\bs{n}}\bra{\bs{n}}U(T,1)\Bigr)^\dagger\nonumber \\ 
&= (\mathds{1}\otimes\mathds{1})^\dagger=\sum_{\bs{mn}}\bbra{\bs{m}\bs{m}\bs{n}\bs{n}}= \sum_{\bs{mn}}\bbra{\bs{m}\bs{n}\bs{n}\bs{m}}=\bbra{\pone},
    \end{align}
where we have exchanged copies $2$ and $4$.
Hence, we write the fidelity as 
\begin{equation}
    \mathcal{F}=\frac{1}{d}\bbra{\pone}\prod^{T-1}_{\tau=0}\Bigl((1-\kappa)\mathds{1}+\kappa \bu^\dagger(T-\tau,1)\bk_{T-\tau}\bu(T-1-\tau,1)\Bigr)\kket{\pthree}
\end{equation}

We are now ready to take the average of the quantity above, and it is easy to see that it involves products of the object
\begin{align}
    \tilde{\bk}_{T-\tau}&\equiv \overline{\bu^\dagger(T-\tau,1)\bk_{T-\tau}\bu(T-1-\tau,1)}
    \nonumber\\
    &=\sum^r_{\mu_{T-\tau}=1}\overline {U^\dagger(T-\tau,1)K_{\mu_{T-\tau}} U(T-\tau,1)\otimes U(1,T-\tau)K^*_{\mu_{T-\tau}}U^*(T-1-\tau,1)}\nonumber\\
    &\overline{\otimes \ U^\dagger(T-\tau,1)U_{T-\tau} U(T-1-\tau,1)\otimes U^T(T-\tau,1)U^*_{T-\tau} U^*(T-1-\tau,1)}\\\nonumber
    &=\sum^r_{\mu_\tau=1}\overline {U^\dagger(T-\tau,1)K_{\mu_{T-\tau}} U(T-\tau,1)\otimes U(1,T-\tau)K^*_{\mu_{T-\tau}}U^*(T-1-\tau,1)}\otimes \mathds{1}\otimes\mathds{1}.
\end{align}
Since the Kraus operators $K_\mu$ are chopped from a large $rd\times rd$ unitary matrix sampled according to the Haar measure, each of them inherits its properties, in particular, the fundamental left-right invariance.
Because of this, 
\begin{equation}
    \overline {U^\dagger(T-\tau,1)K_{\mu_\tau} U(T-\tau,1)\otimes U(1,T-\tau)K^*_{\mu_\tau}U^*(T-1-\tau,1)}= \overline {K_{\mu_\tau}\otimes K^*_{\mu_\tau}}.
\end{equation}
The Weingarten calculus \cite{weingarten} allows us to compute the average over the two copies of the Kraus operators:
\begin{equation}
    \overline {K_{\mu_\tau}\otimes K^*_{\mu_\tau}}=\frac{1}{dr}\sum_{\bs{m}\bs{n}}\kket{\bs{m}\bs{m}}\bbra{\bs{n}\bs{n}},
\end{equation}
because $(\overline {K_{\mu_\tau}\otimes K^*_{\mu_\tau}})^\tau=1/r^{\tau-1}\overline {K_{\mu_\tau}\otimes K^*_{\mu_\tau}}$, $\tilde{\bk}$ is a projector.
Then, it can be seen that 
\begin{equation}
    \Bigl((1-\kappa)\mathds{1}+\kappa\tilde{\bk}\Bigr)^T= (1-\kappa)^T\mathds{1}+(1-(1-\kappa)^T)\tilde{\bk}.
\end{equation}

Finally, computing the overlaps $\bbrakket{\pone}{\pthree}=d$ and 
\begin{equation}
    \bbra{\pone}\tilde{\bk}\kket{\pthree}=r\frac{1}{dr}\sum_{\bs{m}\bs{m}\bs{m'}\bs{n'}\bs{l}\bs{l'}}\bbra{\bs{mnnm}}\Bigl(\kket{\bs{m'm'l'l'}}\bbra{\bs{n'n'l'l'}}\Bigr)\kket{\bs{llll}}=1,
\end{equation}
we obtain the final result:
\begin{equation}\label{eq: analytical fidelity}
    \mathcal{F}=\frac{1}{d}+(1-\kappa)^T\left(1-\frac{1}{d}\right).
\end{equation}
Notice that this expression only depends on the dissipation parameter $\kappa$, indicating that neither the rank $r$ nor the structure of the clean circuit influences the decay of fidelity.

We now numerically show that the fidelity does not depend on the model considered and that it follows Eq.~\eqref{eq: analytical fidelity}.
Consider a random quantum circuit where in each layer we apply a different diluted unitary, corresponding to a $d^2\times d^2$ matrix, sampled randomly within the same model. The initial state is the faultless pure state $\ket{\Psi_0} = \ket{0}^{\otimes L}$. We measure the fidelity between the ideal closed evolution with $\kappa = 0$ and an imperfect map realization with $\kappa \neq 0$. The average fidelity is computed for several values of $\kappa$ and $r$.

\begin{figure}[tbp]
    \centering
    \includegraphics[width=0.6\linewidth]{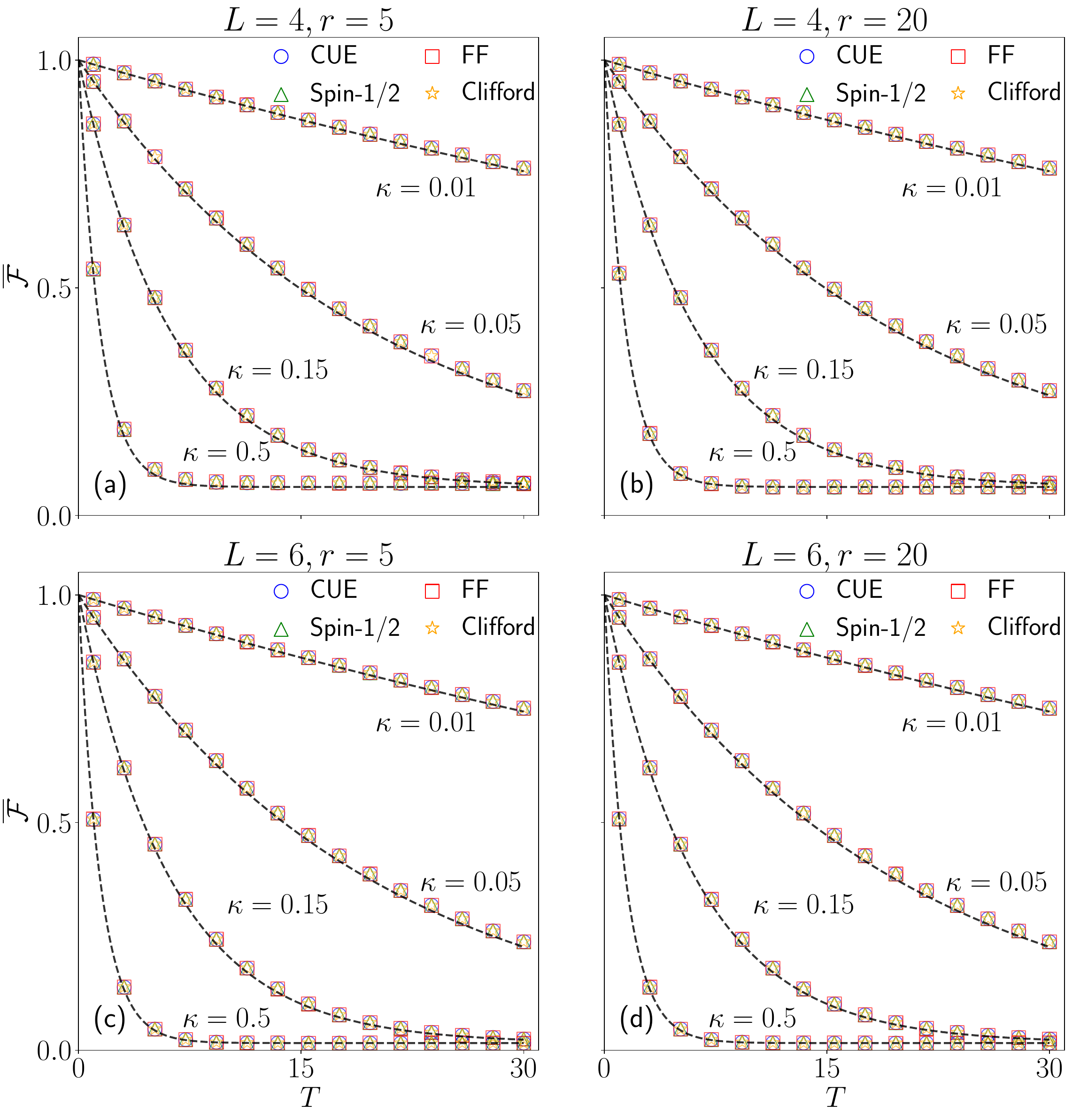}
    \caption{Fidelity decay after applying faulty quantum maps considering different models for different values of $\kappa$, system size, and Kraus rank (see caption for the values). The dashed line represents the fidelity decay of Eq.~\eqref{eq: analytical fidelity}.}
    \label{fig: Fidelity expressibilities}
\end{figure}

In Fig.~\ref{fig: Fidelity expressibilities} \textbf{(a)} we plot the average fidelity as a function of the number of layers $T$ for the four classes of maps considered in the main text, with $\kappa = 0.01, 0.05, 0.15, 0.5$, fixed rank $r = 5$, and $L = 4$. In Fig.~\ref{fig: Fidelity expressibilities} \textbf{(b)} we repeat the process but now with $r = 20$. Finally, Figs.~\ref{fig: Fidelity expressibilities} \textbf{(c)} and \textbf{(d)} represent the same as Figs.~\ref{fig: Fidelity expressibilities} \textbf{(a)} and \textbf{(b)}, respectively, but with $L = 6$.
From Figs.~\ref{fig: Fidelity expressibilities} \textbf{(a)} and \textbf{(b)}, we observe that the fidelity decay is the same for the four maps considered, even for small values of $\kappa$ and $r$, and is well captured by Eq.~\eqref{eq: analytical fidelity}.
This is surprising not only because we cannot distinguish between the behavior of the four maps generated by non-universal gates, but also because their fidelity is the same as in the case of unitaries belonging to the CUE, which is a random ensemble that generates generic dynamics. Furthermore, the rank measures the number of ways the environment interacts with the system; hence, this result is even more peculiar since we observe that, no matter how the environment perturbs the system, the effects on the fidelity will be, on average, the same. Finally, the results from Figs.~\ref{fig: Fidelity expressibilities} \textbf{(c)} and \textbf{(d)} show that the equality of fidelity decay is also observed for other system sizes. 
\par
It is thus safe to conclude that fidelity decay is not a good measure for distinguishing the degree of non-unitarity of different maps.

\section{Robustness of ring-to-disk transition}\label{app: ring to disk}

In this section, we provide numerical evidence of the ring-to-disk robustness that motivated some of the results in the main text.

In the main text, we argued that integrable models with degeneracies undergo two transitions, a cluster-to-ring (C-R) followed by a ring-to-disk (R-D). This was observed through the spectrum of maps with unitaries taken from different ensembles. However, it was also mentioned that before the C-R transition, the absolute value of the eigenvalues would have minimum and maximum radii approximately given by Eq.~(\ref{eq: radii}).

\begin{figure}[tbp]
    \centering
    \includegraphics[width=0.6\linewidth]{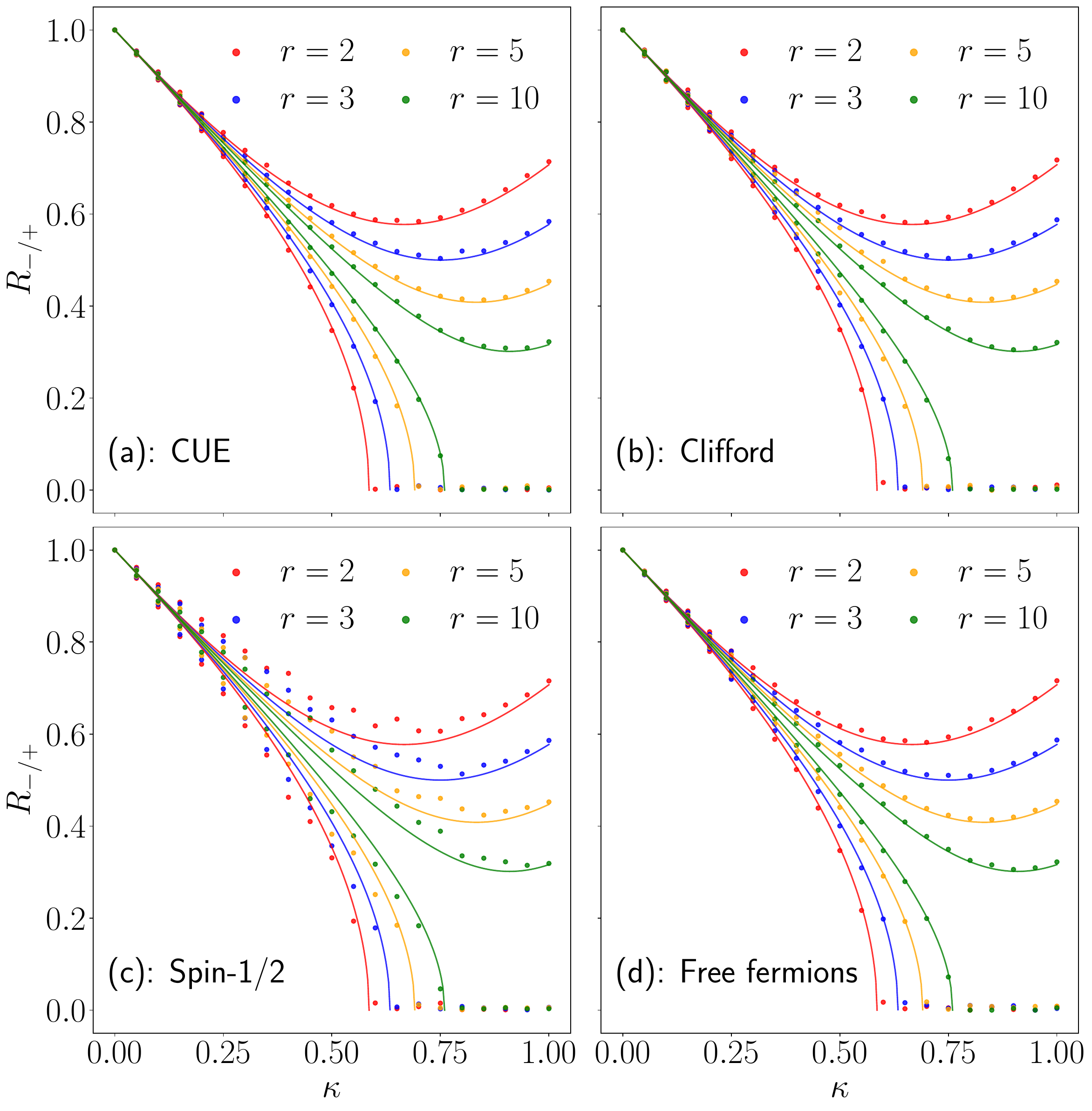}
    \caption{Inner and outer radius, $R_-$ and $R_+$, of the quantum map Eq.~(\ref{eq: five maps}) with a unitary drawn from the CUE \textbf{(a)}, Clifford group \textbf{(b)}, Integrable spin-1/2 \textbf{(c)}, and free fermions ensemble \textbf{(d)} for $L = 6$ and different $r$. The dots represent numerical points obtained with a random map sampled each time for distinct $r$, and the lines the theoretical prediction of Eq.~\eqref{eq: radii}.}
    \label{fig: Rpm}
\end{figure}

In Figs.~\ref{fig: Rpm} \textbf{(a)}--\textbf{(d)} we show that Eq.~\eqref{eq: radii} is valid for the models we considered.
The data points were obtained by sampling a random map for each value of $r$. 
We see that even considering a single map, the data matches well the prediction. The case of the spin-$1/2$ chain seems to be the one that deviates the most from the prediction. However, by inspecting the spectrum of Fig.~\ref{fig: spectrum}, we see the formation of a cluster in the vicinity of the positive real axis, which can lead to some points having an absolute value larger than $R_+$. Nevertheless, the prediction for $R_-$ matches the expected one very well, especially for $\kappa \approx \kappa_{\text{R-D}}$.
\par
These results have two major implications: first, it means that the R-D transition value is approximately the same for all the models; second, it is approximately true that the radius of the clusters is $R = (R_+-R_-)/2$, and that their centers are in the circumference of radius $D = (R_++R_-)/2$, as assumed in the main text.

\section{Method to estimate the number of clusters}
\label{app: cluster count}

In this section, we provide a more detailed description of the method mentioned in the main text to estimate the number of clusters of a map. We apply this method to the spectra depicted in Fig.~\ref{fig: spectrum} of the main text.
\par

Starting with a random unitary from the Clifford group, we notice that the degeneracies are exact. Thus, to count the number of clusters, we simply count the number of distinct eigenvalues. Since the eigenvalues of a Clifford unitary can always be represented geometrically by a complex root of $1$, they form part of a regular polygon with vertices in the unitary circle. As such, a unitary sampled from the Clifford ensemble $U_{\text{Clif}}$ and the corresponding map $\Phi_M = U_{\text{Clif}} \otimes U_{\text{Clif}}^*$ have the same distinct eigenvalues. Therefore, in this case, we consider only the $2^L \times 2^L$ unitary matrix. This is more scalable than the method that follows, which requires the spectrum of the entire map, corresponding to a $4^L \times 4^L$ matrix.

In the other cases, to obtain the number of clusters, we require the unitary map $\Phi_M = U_M \otimes U_M^*$ (or one with a low dissipation strength), where the subscript $M$ indicates the ensemble from which the unitary is sampled. We obtain the spectrum of the map in terms of the angles $\theta$, and for each eigenvalue $\theta_k$, we associate a Gaussian distribution centered at $\theta = \theta_k$
\begin{equation*}
    \rho_k(\theta) = \dfrac{1}{s\cdot\Delta\sqrt{2\pi}} \exp{-\frac{1}{2}\left(\frac{\theta - \theta_k}{s\cdot\Delta}\right)^2},
\end{equation*}
where $\Delta$ is the angular mean level spacing, and $s$ is a scaling factor that allows control over the width of the Gaussian. The quantity of interest is the probability density distribution
\begin{equation}\label{eq: density of points}
    \rho(\theta) = \frac{1}{4^L} \sum_{k = 1}^{4^L} \rho_k(\theta).
\end{equation}
In regions where the density of points is high, we expect Eq.~\eqref{eq: density of points} to reach a local maximum, and a local minimum when the density of points is low. Thus, the idea of this method consists of computing Eq.~\eqref{eq: density of points} associated with the unitary map and counting the number of local maxima, which corresponds to the number of clusters.

\par
Here, the scaling factor $s$ plays a significant role. The angular mean level spacing does not resolve the cluster separation, and we need to widen the width of the Gaussian. We expect that, when the density function becomes smooth around the maxima, the number of clusters will reach a plateau. Therefore, we calculate Eq.~\eqref{eq: density of points} and extract the number of clusters associated with the function with increasing $s$. When we reach a significant plateau, we consider its value as the estimated number of clusters.

\par
\begin{figure}[tbp]
    \centering
    \includegraphics[width=0.8\linewidth]{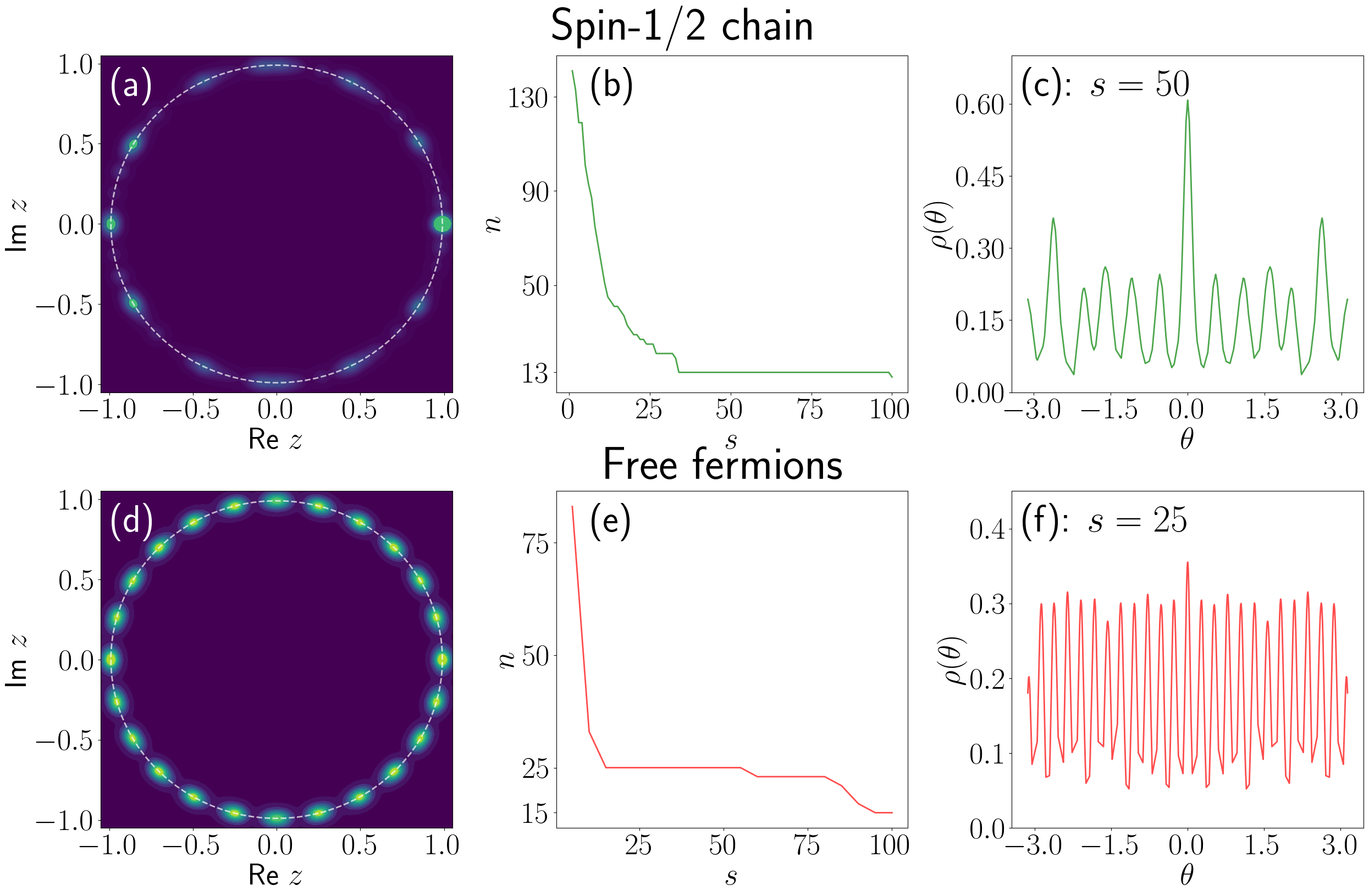}
    \caption{\textbf{(a)}: Spectrum of a random unitary map sampled from the free fermions ensemble. \textbf{(b)}-\textbf{(d)}: Probability density function of Eq.~\eqref{eq: density of points} of the corresponding unitary map for different values of the scaling factor $s$.}
    \label{fig: number of clusters}
\end{figure}

In Fig.~\ref{fig: number of clusters} we demonstrate this method using the spectra of Fig.~\ref{fig: spectrum} of the main text. We focus on the integrable spin chain and free-fermion models. Fig.~\ref{fig: number of clusters} \textbf{(a)} shows the spectrum of a map taken from the integrable spin chain ensemble with $\kappa = 0.01$. We compute the number of clusters $n$ as a function of the scaling $s$ for this spectrum in Fig.~\ref{fig: number of clusters} \textbf{(b)}. We observe a plateau at $n = 13$ clusters. However, if we count them from Fig.~\ref{fig: number of clusters} \textbf{(a)} we only observe twelve regions of significant density. This small discrepancy occurs because we count an additional cluster when a cluster is present around the eigenvalue -1. This is due to the discontinuity at $\theta = \pi$. This can be see in the the density function of Fig.~\ref{fig: number of clusters}, where consider clusters at both $\theta = \pm \pi$. However, in general this is not an issue since our overall goal is to obtain an estimation for this quantity.

\par
We repeat the same procedure in the second row of Fig.~\ref{fig: number of clusters}. In Fig.~\ref{fig: number of clusters} \textbf{(d)} we show the spectrum of a map taken from the free-fermion model with $\kappa = 0.01$. In Fig.~\ref{fig: number of clusters} \textbf{(e)} we show the number of clusters as a function to the scaling factor for this particular maps, observing a plateau at $n = 25$, and finally in Fig.~\ref{fig: number of clusters} \textbf{(f)} we show the density function using $s = 25$. Again, there exists a cluster around -1, meaning there is an additional count to the number of clusters estimated.

\par

In practice, the method works well for the spectra considered in this work because the relevant cluster structure is typically associated with a clear plateau as the smoothing parameter is varied. The main source of ambiguity arises only when many clusters lie very close to one another, or when a weak feature near $\theta=\pm\pi$ is split by the angular discontinuity. In such cases the count may differ slightly from an exact visual partition, but it still captures the scale of $n$ that is relevant for the cluster-to-ring transition. Likewise, irregular clusters can occasionally cause the algorithm to resolve a small peak that merges quickly with a larger one as dissipation increases. These effects lead to minor variations in the estimated count, rather than to a failure of the method. For our purposes, the procedure provides a reliable and reproducible estimator of the cluster number entering Eq.~\eqref{eq: transition solution}, and it is sufficiently accurate to identify both the scaling with system size and the relevant crossover regime.

\end{document}